\documentclass[fleqn,usenatbib]{mnras}

\usepackage{newtxtext,newtxmath}

\usepackage[T1]{fontenc}

\DeclareRobustCommand{\VAN}[3]{#2}
\let\VANthebibliography\thebibliography
\def\thebibliography{\DeclareRobustCommand{\VAN}[3]{##3}\VANthebibliography}

\usepackage{graphicx}	
\usepackage{amsmath}	

\newcommand{\msun}{M_\odot}
\newcommand{\msunyr}{M_\odot\,{\rm yr}^{-1}}
\newcommand{\nh}{n_{\rm H}}
\newcommand{\nth}{n_{\rm th}}
\newcommand{\tth}{t_{\rm th}}
\newcommand{\cc}{{\rm cm^{-3}}}
\newcommand{\kms}{{\rm km\,s^{-1}}}

\defcitealias{Hirano2018}{H18}

\title[First star clusters - I]{Formation of first star clusters under the supersonic gas flow - I.\\ Morphology of the massive metal-free gas cloud}

\author[S. Hirano et al.]{
Shingo Hirano,$^{1}$\thanks{E-mail: hirano@astron.s.u-tokyo.ac.jp (SH)}
Youcheng Shen,$^{1}$
Sho Nishijima,$^{1}$
Yusuke Sakai$^{1}$ and 
Hideyuki Umeda$^{1}$
\\
$^{1}$Department of Astronomy, School of Science, University of Tokyo, Tokyo 113-0033, Japan
}

\date{Accepted 2023 September 03. Received 2023 August 28; in original form 2023 June 15}

\pubyear{2023}

\begin{document}
\label{firstpage}
\pagerange{\pageref{firstpage}--\pageref{lastpage}}
\maketitle

\begin{abstract}
We performed $42$ simulations of first star formation with initial supersonic gas flows relative to the dark matter at the cosmic recombination era.
Increasing the initial streaming velocities led to delayed halo formation and increased halo mass, enhancing the mass of the gravitationally shrinking gas cloud.
For more massive gas clouds, the rate of temperature drop during contraction, in other words, the structure asymmetry, becomes more significant.
When the maximum and minimum gas temperature ratios before and after contraction exceed about ten, the asymmetric structure of the gas cloud prevails, inducing fragmentation into multiple dense gas clouds.
We continued our simulations until $10^5$ years after the first dense core formation to examine the final fate of the massive star-forming gas cloud.
Among the $42$ models studied, we find the simultaneous formation of up to four dense gas clouds, with a total mass of about $2254\,\msun$.
While the gas mass in the host halo increases with increasing the initial streaming velocity, the mass of the dense cores does not change significantly.
The star formation efficiency decreases by more than one order of magnitude from $\epsilon_{\rm III} \sim 10^{-2}$ to $10^{-4}$ when the initial streaming velocity, normalised by the root mean square value, increases from 0 to 3.
\end{abstract}



\section{Introduction} \label{sec:intro}

The first stars, Population III (Pop III) stars, formed from the metal-free gas cloud and brought the first light and heavy elements to the universe \citep[see][for a recent review]{Klessen2023}.
The cosmological simulations show that small dark matter (DM) halos of $\sim\!10^5 - 10^6\,\msun$ forming at redshift $z \sim 20 - 30$ become cradles of the first stars \citep[e.g.][]{Tegmark1997,Yoshida2003}.
Typically, a massive star-forming gas cloud of $\sim\!1000\,\msun$ is formed at the density peak of the host halo \citep[e.g.][]{Abel2002,Bromm2002}.
The cloud gravitationally collapses until a quasi-hydrostatic protostellar core of $\sim\!0.01\,\msun$ is formed \citep[e.g.][]{Omukai1998,Yoshida2008}.
The tiny protostellar core grows via accretion of the surrounding gas until the protostellar radiative feedback halts the gas accretion \citep[e.g.][]{McKee2008,Hosokawa2011,Hosokawa2016}.
Previous simulations follow the accretion phase to determine the final stellar mass and construct the initial mass function of Pop III stars \citep[e.g.][]{Hosokawa2011,Hosokawa2016,Hirano2014,Hirano2015}.
Some simulations reported the formation of multiple-star systems \citep[e.g.][]{Susa2013,Susa2014,susa2019,Stacy2013,Stacy2016,Sugimura2020}.

Environmental effects are also known to affect the first star formation process.
Various effects have been investigated using numerical simulations: (dynamical) baryonic supersonic motions relative to DM \citep[e.g.][]{Tseliakhovich2010}, violent halo mergers \citep[e.g.][]{Inayoshi2015,Wise2019}, (radiative) far-ultraviolet radiation in the Lyman-Werner bands \citep[e.g.][]{Omukai2001,Latif2013}, X-ray radiation \citep[e.g.][]{Hummel2005,Park2021}.
In recent years, research has been conducted in a more realistic cosmological setting that deals with these influences simultaneously \citep[e.g.][]{Schauer2021, Kulkarni2021}.
Because these effects support the formation of supermassive clouds by pausing the first star formation, researchers have investigated in the context of supermassive first star formation, which is the candidate of a seed object of supermassive black holes (SMBHs) observed in the distant universe \citep[e.g.][]{Inayoshi2020}.

We focus on the final fate of the massive star-forming clouds formed under the baryonic streaming velocity \citep[SV; e.g.][]{Stacy2011,Greif2011}.
We presented that, in regions with a large SV, gas condensation is suppressed until DM halo generates a deep gravitational potential with a mass of $10^7\,\msun$, and a protostar which is formed in the massive gas cloud and grows its mass via episodically burst accretion \citep{Hirano2017a}.
In regions with low-to-moderate SV, on the other hand, a sizeable filamentary gas cloud of $10^4-10^5\,\msun$ forms and fragments to yield multiple star-forming gas clouds of $100-1000\,\msun$ \citep[][hereafter \citetalias{Hirano2018}]{Hirano2018}.
If each of these cloud clusters forms a multiple first star system, it will form a first star association than the clusters that have been discussed in the past in a single gas cloud.

We aim to identify the formation conditions of the first stars born as a single star, binary, or cluster.
\citetalias{Hirano2018}, however, simulated the first star formation only in the same halo under different initial SVs, despite the known diversity of the star formation \citep[e.g.][]{Hirano2014,Hirano2015}.
To determine the critical initial SV for the multiple cloud formation and the typical number of clouds, we study the statistical properties of massive star-forming clouds in the early universe using $42$ clouds produced by cosmological simulations.
We determine the mass function of the metal-free star-forming clouds at $10^5$\,yr after the first cloud formation by adopting the opaque core methodology \citep{Hirano2017b}.
We find that the rate of temperature drop during contraction determines the morphology of the gas cloud.
The more asymmetric the shape, the more likely it fragments and forms a massive cloud association.

Before moving on to the next section, another first star formation process due to a rapid SV is worth mentioning.
The rapid SV causes a spatial offset between DM and baryon density perturbations, forming baryonic clumps that collapse outside the counterpart DM halos \citep{Naoz2014}.
Such supersonically induced gas objects (SIGOs) could survive as DM-free objects and might become globular clusters \citep[e.g.][]{Popa2016, Chiou2019, Lake2023}.
This study restricts the computational domain within the DM halo to resolve the supermassive clouds by low-mass particles.
We have yet to trace the formation of SIGO outside the DM halo adequately.
Simultaneously investigating the effects of SV on the inside and outside of the halo is a challenge that will require future, larger-scale calculations.

We describe the calculation methods in Section~\ref{sec:method}.
Section~\ref{sec:results} shows the results of the metal-free star-forming gas cloud formation inside $7$ different DM halos under $6$ different initial SVs, $42$ models in total.
Section~\ref{sec:dis} discusses the dependence of first star formation efficiency on the formation environment.
Section~\ref{sec:sum} summarises the parameterised study and provides an outlook for future research.

\section{Numerical Methodology} \label{sec:method}

We perform a set of cosmological simulations under different initial baryonic streaming velocities to study the effect of the early streaming motions on the first star formation.
We calculate the first $10^5$\,yr evolution of the star-forming region after the first cloud formation ($\sim$ protostar formation epoch) to examine the cloud formation until the protostellar radiative feedback from massive first stars evaporates the accreting gas material.
We discuss the effect of the initial streaming velocities on the physical properties of star-forming clouds.

\subsection{Cosmological Initial Conditions} \label{sec:method:ics}

\begin{figure}
\begin{center}
\includegraphics[width=0.9\linewidth]{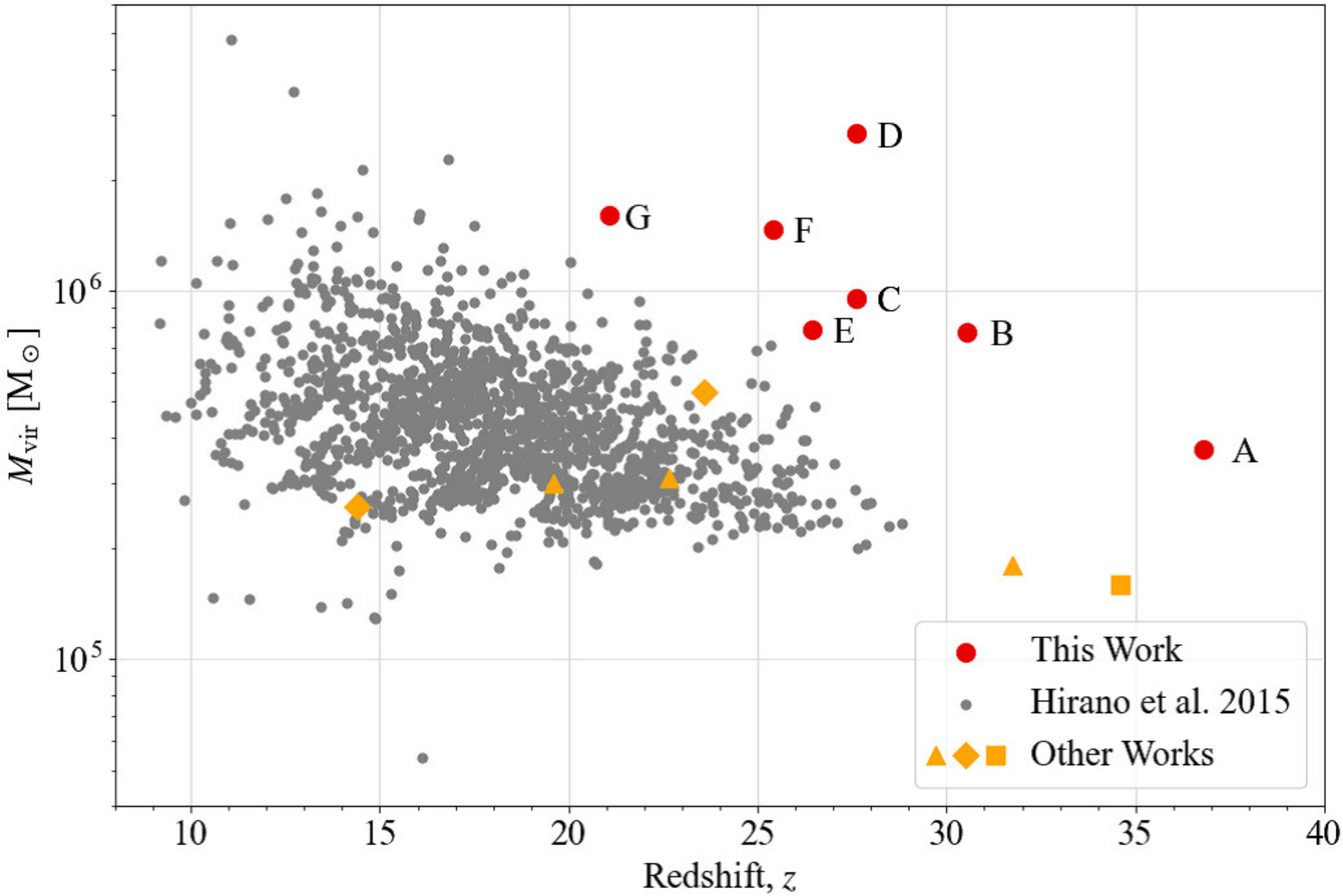}
\end{center}
\caption{
Redshift and virial halo mass distribution obtained from the cosmological simulations with zero streaming velocity.
The red circles indicate the seven target halos in this study (Halos A-G).
The grey dots, yellow triangles, diamond, and square indicate models in the previous works, \citet{Hirano2015}, \citet{Stacy2011}, \citet{Greif2011}, and \citetalias{Hirano2018}, respectively.
}
\label{f1}
\end{figure}

We use the publicly available code \texttt{MUSIC} \citep{Hahn2011} to generate the cosmological initial conditions (ICs) with a volume of $L_{\rm box} = 10\,h^{-1}$ comoving megaparsec (cMpc) on a side at redshift $z_{\rm ini} = 499$.
We adopt the standard $\Lambda$-cold DM ($\Lambda$CDM) cosmology with 
total matter density $\Omega_{\rm m} = 0.31$, 
baryon density, $\Omega_{\rm b} = 0.048$, 
dark energy density $\Omega_\Lambda = 0.69$ in units of the critical density, 
Hubble constant $h = 0.68$, density fluctuation amplitude $\sigma_8 = 0.83$, and 
primordial spectral index $n_{\rm s} = 0.96$.

We perform a set of cosmological simulations with two parameters: 
(1) DM halo which hosts the metal-free star-forming cloud and 
(2) initial velocities of the baryonic streaming motion.
First, we simulate seven cosmological simulations with zero streaming velocity (SV) initiated by seven cosmological ICs.
We identify the virial DM minihalos first formed in each simulation volume.
We label these target halos as Halos A-G in order of the formation redshift ($z = 21.08 - 36.80$).
This is the first parameter in this study.
For comparison, Figure~\ref{f1} overplots the redshift-halo mass diagram of our target halos on the scatter plots of the samples examined in the previous studies \citep[][\citetalias{Hirano2018}]{Stacy2011, Greif2011, Hirano2015}.
Halos~A-G are more massive and form earlier than the previous samples since we select the first formed halo in a larger cosmological box.
Because SV decreases with time as $v_{\rm SV} \propto (1+z)$, the influence of SV increases at a higher redshift.

Next, we generate the cosmological ICs (labels as A-G) by adding a uniform initial relative velocity between DM and baryonic components along the $x$-axis.
We adopt a uniform initial relative velocity since the distribution of the baryonic streaming motion is coherent over a length of a few megaparsecs, which is sufficiently larger than a typical scale that contains target DM haloes.
We select six different initial SVs to study the dependence: $v_{\rm SV} / \sigma_{\rm SV}^{\rm rec} = 0$, $1.0$, $1.5$, $2.0$, $2.5$, and $3.0$, normalised by the root-mean-square value, $\sigma_{\rm SV}^{\rm rec} = 30\,\kms$, at the epoch of the cosmological recombination, $z_{\rm rec} = 1089$.\footnote{\citet{UysalHartwig2023} estimated the local value of SV in which the Milky Way was formed as $v_{\rm SV}/\sigma_{\rm SV}^{\rm rec} = 1.75$, which is an extremely high value, and if valid, its impact on the structure formation in the early universe is inescapable.}
The probability fraction of the cosmological volume where the streaming velocities $v_{\rm SV} / \sigma_{\rm SV}^{\rm rec} = 1$, $2$, and $3$ are $0.39$, $7.4 \times 10^{-3}$, and $5.9 \times 10^{-6}$, respectively \citep{Tseliakhovich2011}.
Then the range of the initial SV in this study covers the possible formation environment.
This study investigates $7 \times 6 = 42$ models in total (Table~\ref{t1}).
Hereafter, we refer to each model by a combination of one capital alphabet letter and two numbers; for example, C15 represents Halo C with $v_{\rm SV} / \sigma_{\rm SV}^{\rm rec} = 1.5$.

We regenerate the cosmological ICs with a hierarchical zoom-in region with a volume of $L_{\rm zoom} = 0.3\,h^{-1}$\,cMpc on a side.
In the high-resolution regions, the particle masses of DM and gas components are $m_{\rm DM} = 16.4\,\msun$ and $m_{\rm gas} = 3.0\,\msun$, respectively.
This particle mass is enough to resolve the host DM halo with $>\!10^6\,\msun$.

\begin{table*}
\centering
\caption{
Column 1: halo name.
Column 2: relative streaming velocity normalised by the root-mean-square value $v_{\rm sv}/\sigma_{\rm sv}$.
Column 3: redshift $z$ when the gas number density firstly reaches $\nh = 10^6\,\cc$.
Columns 4 to 8: radius $R_{\rm v}$, total mass $M_{\rm v}$, gas temperature $T_{\rm v}$, gas mass $M_{\rm v,b}$, and gas mass fraction $f_{\rm b}$ at the virial scale.
Column 9: BE mass of the cloud $M_{\rm J,whole}$ when it first becomes gravitationally unstable. 
Column 10: classification of the gas cloud structure. 
Columns 11 and 12: numbers of clouds $N_{\rm 6}$ (for $\nh > 10^6\,\cc$) and cores $N_{\rm 8}$ (for $\nh > 10^8\,\cc$), respectively.
Column 13: BE mass of the cloud $M_{\rm J}$ when it becomes the most gravitationally unstable. 
Column 14: core mass $M_{\rm core}$ where $\nh \geq 10^8\,\cc$.
For models with multiple cores, data for each core are noted in the 13 and 14th columns.
Pairs of cores that came within the Jeans radius ($r_{\rm J} = 0.25$\,pc) of each other during the calculation are marked with asterisks (* and **) in the 13th column.
}
\label{t1}
\begin{tabular}{@{}clcrlrlcrcccrr@{}} 
\hline
Halo & $v_{\rm sv}/\sigma_{\rm sv}$ & $z$ & $R_{\rm v}$ & $M_{\rm v}$ & $T_{\rm v}$ & $M_{\rm v,b}$ & $f_{\rm b}$ & $M_{\rm J,whole}$ & Type & $N_{\rm 6}$ & $N_{\rm 8}$ & $M_{\rm J}$ & $M_{\rm core}$ \\
& & &$(\rm pc)$ &$(10^7\,\msun)$ &(K) &$(10^6\,\msun)$ & &$(\msun)$ & & & &$(\msun)$ &$(\msun)$ \\
\hline
A & 0   & 36.80 &  63 & 0.037 &  280 & 0.040 & 0.109 & $6.0\times10^3$ & S &   &   &    1068 &  516 \\ 
  & 1.0 & 30.91 & 126 & 0.19  & 1014 & 0.201 & 0.106 & $4.0\times10^3$ & S &   &   &    1559 & 1316 \\ 
  & 1.5 & 27.45 & 224 & 0.79  & 1380 & 0.901 & 0.114 & $1.1\times10^4$ & F &   &   &     652 &  842 \\ 
  & 2.0 & 25.84 & 316 & 1.79  &  582 & 2.11  & 0.118 & $4.0\times10^4$ & F & 2 &   &     462 & 1083 \\ 
  & 2.5 & 25.10 & 355 & 2.42  & 1566 & 2.63  & 0.109 & $2.5\times10^4$ & F & 2 & 2 &     502 & 1076 \\ 
  &     &       &     &       &      &       &       &                 &   &   &   &     508 & 1663 \\ 
  & 3.0 & 23.33 & 562 & 6.47  & 4189 & 7.83  & 0.121 & $3.2\times10^4$ & F & 2 & 2 &     458 & 1628 \\ 
  &     &       &     &       &      &       &       &                 &   &   &   &     355 &  752 \\ 
  \hline
B & 0   & 30.53 & 100 & 0.077 &  647 & 0.089 & 0.116 & $2.5\times10^3$ & S &   &   &     654 &  280 \\ 
  & 1.0 & 30.22 & 100 & 0.08  &  709 & 0.091 & 0.114 & $1.2\times10^3$ & S &   &   &     496 &  268 \\ 
  & 1.5 & 27.45 & 141 & 0.16  &  897 & 0.204 & 0.128 & $7.0\times10^3$ & S & 2 &   &    2123 &  109 \\ 
  & 2.0 & 24.82 & 251 & 0.68  & 1159 & 0.741 & 0.109 & $2.0\times10^4$ & F &   &   &    1480 &  578 \\ 
  & 2.5 & 23.02 & 316 & 1.18  & 1636 & 1.33  & 0.113 & $4.0\times10^4$ & F &   &   &    1199 & 1278 \\ 
  & 3.0 & 23.31 & 282 & 0.88  & 1597 & 0.783 & 0.089 & $4.5\times10^4$ & F & 2 &   &    4554 & 1247 \\ 
  \hline
C & 0   & 27.61 & 112 & 0.095 &  873 & 0.135 & 0.143 & $1.3\times10^3$ & S & 3 & 2 & $^*$242 &  234 \\ 
  &     &       &     &       &      &       &       &                 &   &   &   & $^*$273 &  107 \\ 
  & 1.0 & 25.38 & 178 & 0.26  &  879 & 0.358 & 0.138 & $5.0\times10^3$ & F &   &   &    1225 &  585 \\ 
  & 1.5 & 24.20 & 200 & 0.38  & 1038 & 0.494 & 0.130 & $6.0\times10^3$ & F &   &   &    1079 &  793 \\ 
  & 2.0 & 23.04 & 251 & 0.68  & 1243 & 0.897 & 0.132 & $6.0\times10^3$ & F &   &   &    1599 &  689 \\ 
  & 2.5 & 21.37 & 398 & 2.01  & 1634 & 2.97  & 0.148 & $6.0\times10^4$ & C &   &   &    1137 &  843 \\ 
  & 3.0 & 20.79 & 447 & 2.72  & 1454 & 3.64  & 0.134 & $1.6\times10^5$ & C & 5 &   &    2222 & 1361 \\ 
  \hline
D & 0   & 27.60 & 158 & 0.27  & 1103 & 0.342 & 0.127 & $2.7\times10^3$ & S &   &   &     854 & 1645 \\ 
  & 1.0 & 25.42 & 224 & 0.59  & 1124 & 0.826 & 0.140 & $3.5\times10^4$ & F &   &   &     666 &  654 \\ 
  & 1.5 & 24.51 & 282 & 1.01  & 1980 & 1.26  & 0.125 & $4.4\times10^4$ & F &   &   &    2238 &  849 \\ 
  & 2.0 & 23.75 & 316 & 1.34  & 1648 & 1.55  & 0.116 & $2.8\times10^4$ & F &   &   &    1704 &  756 \\ 
  & 2.5 & 21.09 & 501 & 4.03  & 3412 & 4.63  & 0.115 & $6.0\times10^4$ & C & 6 & 4 &     603 &  792 \\ 
  &     &       &     &       &      &       &       &                 &   &   &   &     785 &  782 \\ 
  &     &       &     &       &      &       &       &                 &   &   &   &     692 &  628 \\ 
  &     &       &     &       &      &       &       &                 &   &   &   &     560 &   52 \\ 
  & 3.0 & 20.85 & 501 & 4.06  & 3430 & 4.38  & 0.108 & $1.0\times10^5$ & C & 4 & 3 &     891 & 1094 \\ 
  &     &       &     &       &      &       &       &                 &   &   &   & $^{**}$683 &  759 \\ 
  &     &       &     &       &      &       &       &                 &   &   &   & $^{**}$492 &  524 \\ 
  \hline
E & 0   & 26.43 & 112 & 0.078 &  602 & 0.110 & 0.141 & $1.2\times10^3$ & S &   &   &     172 &  970 \\ 
  & 1.0 & 23.46 & 178 & 0.23  &  586 & 0.331 & 0.144 & $3.0\times10^3$ & S &   &   &     408 & 1009 \\ 
  & 1.5 & 21.70 & 251 & 0.45  & 1081 & 0.531 & 0.118 & $1.3\times10^4$ & F & 2 &   &     731 & 1214 \\ 
  & 2.0 & 19.55 & 316 & 0.78  & 1075 & 0.873 & 0.112 & $2.0\times10^4$ & F &   &   &     378 & 1390 \\ 
  & 2.5 & 17.93 & 398 & 1.26  & 1212 & 1.49  & 0.119 & $7.0\times10^4$ & C &   &   &     452 & 1007 \\ 
  & 3.0 & 17.23 & 447 & 1.65  & 1564 & 1.88  & 0.114 & $8.5\times10^4$ & C &   &   &     247 &  663 \\ 
  \hline
F & 0   & 25.42 & 141 & 0.15  & 1025 & 0.222 & 0.148 & $8.0\times10^3$ & S &   &   &    1403 &  182 \\ 
  & 1.0 & 22.70 & 224 & 0.46  & 1190 & 0.634 & 0.138 & $1.5\times10^4$ & C &   &   &     642 & 1186 \\ 
  & 1.5 & 22.21 & 251 & 0.56  &  944 & 0.711 & 0.127 & $1.6\times10^4$ & F &   &   &     702 & 1014 \\ 
  & 2.0 & 21.27 & 282 & 0.75  & 1532 & 0.900 & 0.120 & $3.6\times10^4$ & F &   &   &    1435 & 1122 \\ 
  & 2.5 & 20.50 & 355 & 1.13  & 1313 & 1.35  & 0.120 & $3.0\times10^5$ & F &   &   &    1071 & 1374 \\ 
  & 3.0 & 19.79 & 398 & 1.47  & 1227 & 1.82  & 0.124 & $5.5\times10^4$ & C &   &   &    1214 &  363 \\ 
  \hline
G & 0   & 21.08 & 178 & 0.16  & 1060 & 0.235 & 0.147 & $2.0\times10^3$ & S &   &   &     272 & 1019 \\ 
  & 1.0 & 20.98 & 178 & 0.16  &  718 & 0.233 & 0.146 & $3.8\times10^3$ & S &   &   &     166 &  526 \\ 
  & 1.5 & 19.19 & 224 & 0.27  &  986 & 0.429 & 0.159 & $4.0\times10^3$ & S &   &   &     139 &  317 \\ 
  & 2.0 & 17.33 & 398 & 0.97  & 1150 & 1.41  & 0.146 & $2.0\times10^4$ & C & 2 &   &     727 &  162 \\ 
  & 2.5 & 17.01 & 398 & 1.04  & 1314 & 1.42  & 0.137 & $9.0\times10^4$ & S &   &   &     262 &  858 \\ 
  & 3.0 & 16.00 & 447 & 1.36  & 1538 & 1.86  & 0.137 & $3.8\times10^4$ & C &   &   &     619 &  281 \\ 
\hline
\end{tabular}
\end{table*}

\begin{figure*}
\begin{center}
\includegraphics[width=0.9\linewidth]{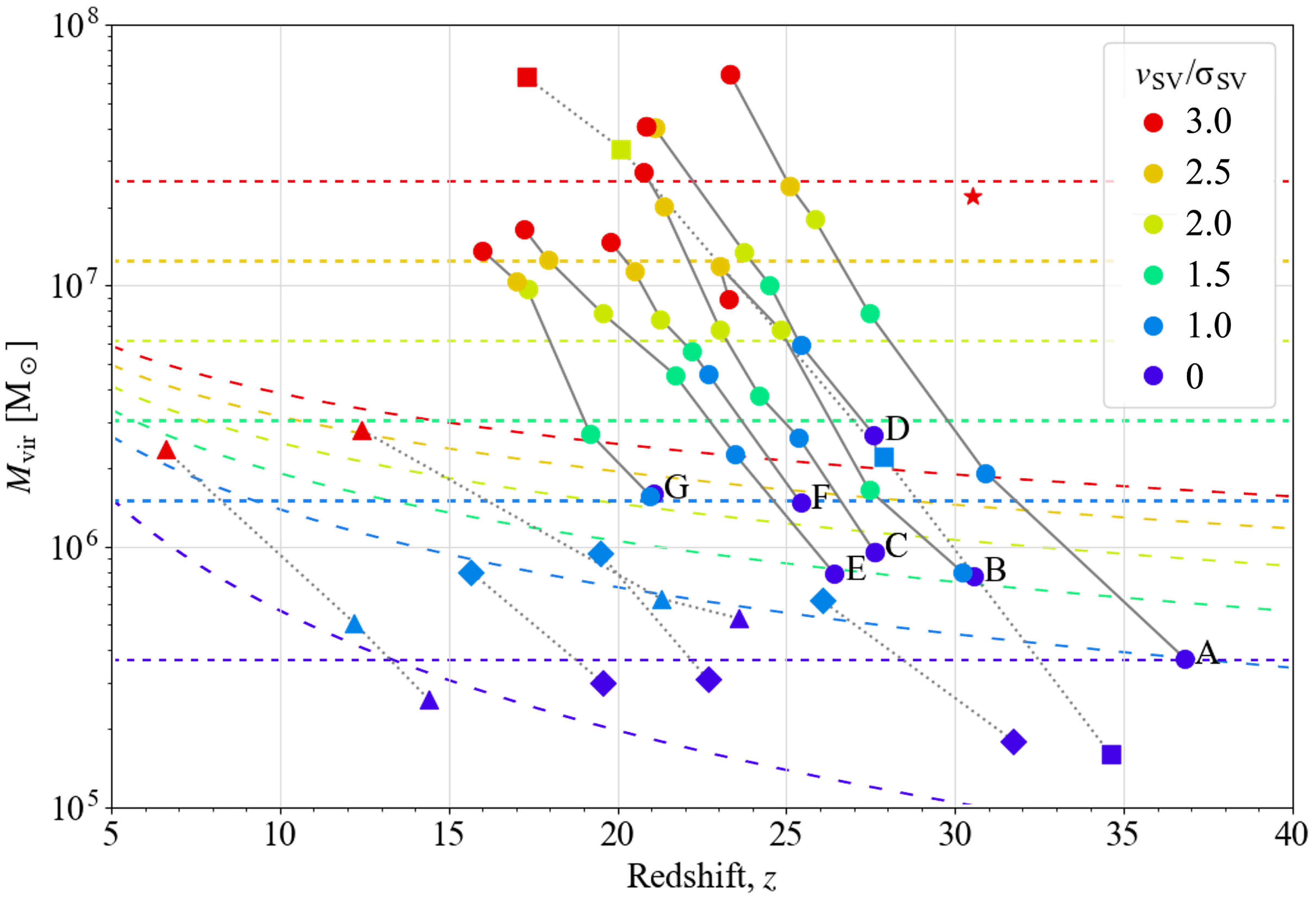}
\end{center}
\caption{
Distribution of redshift and virial halo mass.
The circles indicate $42$ models in this study.
The triangles, diamonds, stars, and squares indicate models in the previous works, \citet{Stacy2011}, \citet{Greif2011}, \citet{Hirano2017a}, and \citetalias{Hirano2018}, respectively.
The horizontal dotted and curving dashed lines show fitting functions of the minimum halo mass in \citet[][Equations~11-13]{Schauer2021} and \citet[][Equations~2-12]{Kulkarni2021} with no Lyman-Werner background.
The lines connect models initiated by the same cosmological region by adding different initial streaming velocities.
As shown in the legend, the colours of symbols and lines correspond to the magnitude of the initial streaming velocity, except for the red triangle where $v_{\rm SV}/\sigma_{\rm SV} = 3.3$ \citep{Stacy2011}.
}
\label{fig:z-Mvir}
\end{figure*}

\begin{figure}
\begin{center}
\includegraphics[width=1.0\columnwidth]{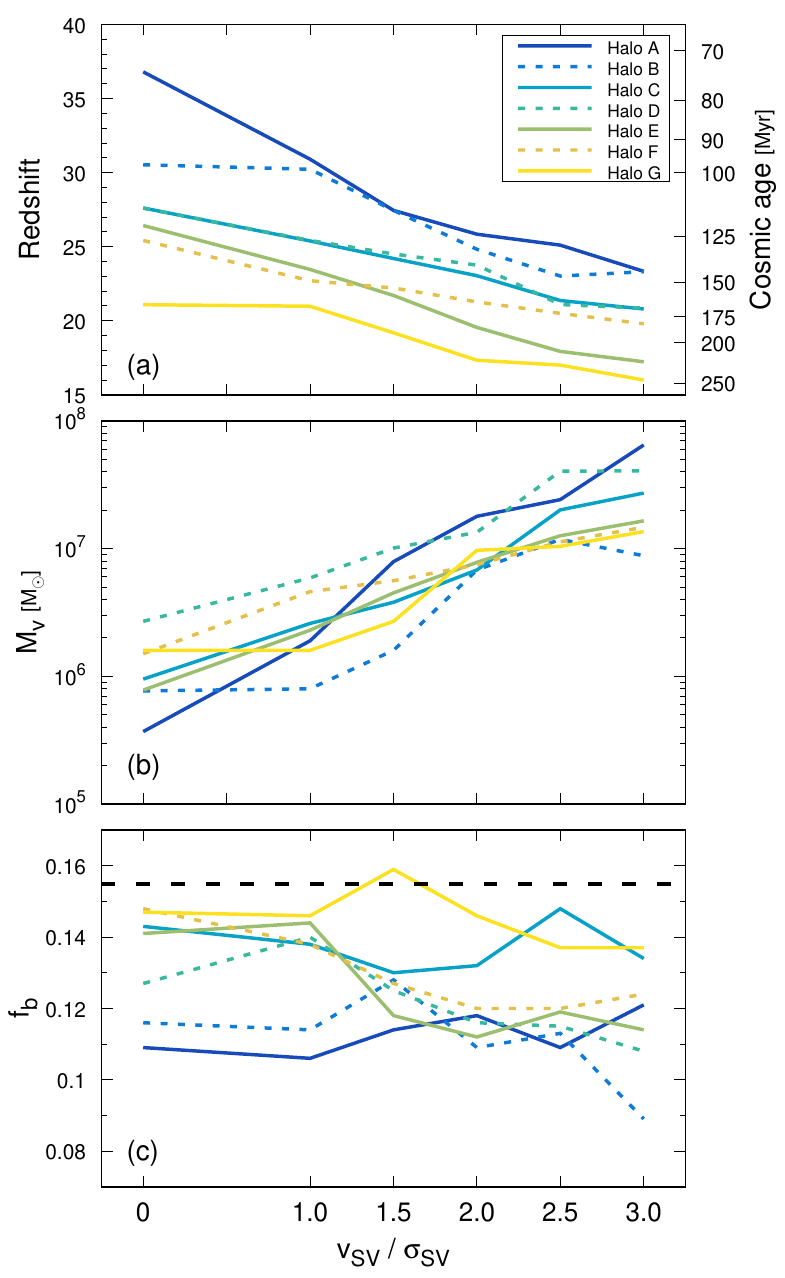}
\end{center}
\caption{
Physical properties of the virial halos as a function of the initial streaming velocity.
Panels: (a) formation redshift and corresponding cosmic age, (b) virial halo mass, and (c) baryon fraction, respectively.
The horizontal dashed line in panel (c) indicates the cosmological mean baryon fraction, $\Omega_{\rm b} / \Omega_{\rm m} = 0.155$.
}
\label{fig:Vsv-Halo}
\end{figure}

\begin{figure}
\begin{center}
\includegraphics[width=1.0\columnwidth]{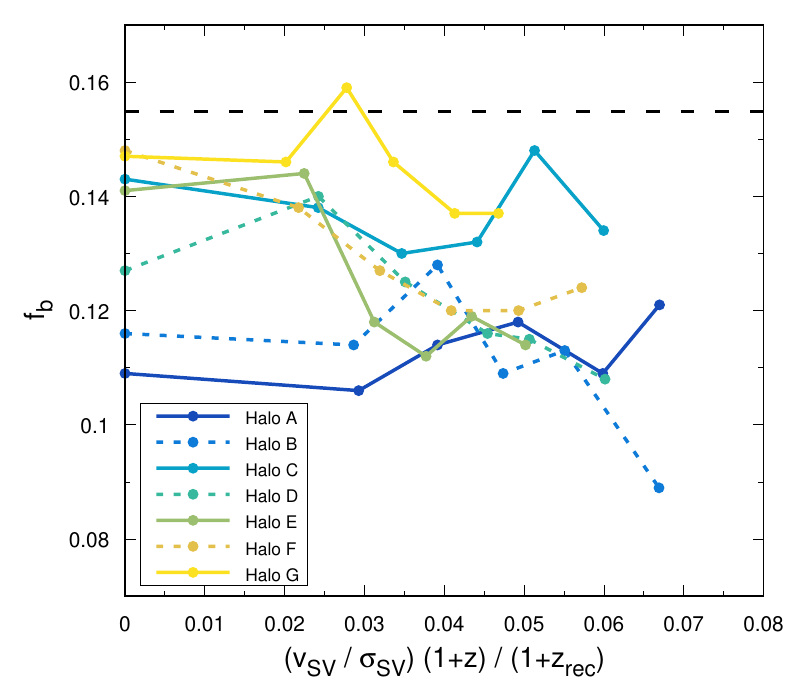}
\end{center}
\caption{
Baryon fraction of the viral halo as a function of the streaming velocity at the time of the collapse, $(v_{\rm SV}/\sigma_{SV})(1+z)/(1+z_{\rm rec})$.
The horizontal dashed line indicates the cosmological mean baryon fraction, $\Omega_{\rm b} / \Omega_{\rm m} = 0.155$.
}
\label{fig:Vsvcol-fb}
\end{figure}

\subsection{Cosmological Simulations} \label{sec:method:cosmosim}

We perform a set of cosmological simulations using the parallel $N$-body/smoothed particle hydrodynamics (SPH) code \texttt{GADGET-2} \citep{Springel2005} suitably adopted for the metal-free star formation (\citetalias{Hirano2018}).
We solve chemical reactions for $14$ species in the primordial gas (e$^-$, H, H$^+$, H$^-$, He, He$^+$, He$^{++}$, H$_2$, H$^+_2$, D, D$^+$, HD, HD$^+$, and HD$^-$) as in \citep{Yoshida2007,Yoshida2008}.
We use the updated cooling rates for H$_2$ and HD \citep{GalliPalla2013} and the three-body H$_2$ formation rates \citep{Forrey2013a, Forrey2013b}.

We employ a hierarchical refinement technique to follow the first star formation process.
We adopt the refinement criterion that the local Jeans length of the SPH particle is always resolved by $15$ times the local smoothing length by increasing the spatial resolution using the particle-splitting technique \citep{Kitsionas2002}, which places the 13 child particles on a hexagonal close-packed array.\footnote{\citet{ChiakiYoshida2015} pointed out that the particle splitting method, which places the particles spherically symmetrical, could wipe out the original non-spherically symmetrical density structure. This study adopted a stricter condition as $R_{\rm cr}=M_{\rm Jeans}/m=15^3$ than $R_{\rm cr}=10^3$ in our previous simulations to complete the particle splitting before forming small density structures, thereby reducing the influence of particle splitting methods that assume spherically symmetric structures on non-spherically symmetric density structures. In addition, this study focuses on cloud-scale fragmentation, which occurred in lower dense and more spherical region than disk-scale fragmentation discussed in \citet{ChiakiYoshida2015}. Therefore, the influence of using the spherical symmetry particle splitting method is considered small.}

The simulations end when the hydrogen number density, $\nh = \rho/m_{\rm H}$, first reaches $10^8\,\cc$.
The minimum mass of the gas particle is $m_{\rm gas} = 0.025\,\msun$ in this study.

\subsection{Hydrodynamic Simulations} \label{sec:method:hydrosim}

The cosmological simulations end before the protostar formation.
To examine the formation and long-term evolution of the star-forming cloud, we rerun all models using an opaque core methodology \citep{Hirano2017b}.
Our method artificially reduces the radiative cooling for gas particles whose density exceeds a threshold value, $\nth = 10^8\,\cc$, as
\begin{equation}
  \Lambda_{\rm red} = \beta_{\rm esc,art} \cdot \Lambda_{\rm thin} \, ,
  \label{eq:Lambda}
\end{equation}
with an artificial escape fraction and an artificial optical depth as
\begin{equation}
  \beta_{\rm esc,art} = \frac{1-\exp(\tau_{\rm art})}{\tau_{\rm art}}, \tau_{\rm art} = \left(\frac{n}{\nth}\right)^2 \, .
  \label{eq:EscapeFraction}
\end{equation}
Dense regions exceeding the threshold density are experiencing compression heating and forming a hydrostatic core.
In addition, we skip the calculations of chemical reactions for gas particles whose density exceeds the threshold value.
We clarify that the radius of the dense core is less than $0.05$\,pc in all models, which does not affect the Jeans-scale structure analysed later.

We stop all runs with the opaque core methodology $10^5$\,yr after the gas particles reach the threshold density ($\nth = 10^8\,\cc$ at $\tth = 0$\,yr).
This calculation time is sufficiently longer than the free-fall time at the threshold density, $t_{\rm ff} = 5.2 \times 10^3 (\nh/10^8\,\cc)^{-1/2}$\,yr, so the protostar can form inside the opaque core in this time.
The newly born protostar evolves to the zero-age main sequence phase on average $\sim\!10^5$\,yr in the case of the first star formation \citep[see Figure~1 in][]{Hirano2017b} and begins to blow off the surrounding gas due to the UV radiative feedback \citep{McKee2008}.
We end the calculations in this study at this time because our simulations ignore the UV radiative feedback.

\subsection{Cloud and Core} \label{sec:method:host}

We analyse the time-series data from the long-term simulations to determine the number of high-density regions inside which the first star can form.
We define dense regions at two scales using different critical densities:
(1) $\nh = 10^6\,\cc$ above which the collapsing cloud is already gravitationally unstable and
(2) $\nh = 10^8\,\cc$ above which the opaque core methodology suppresses the gravitational collapse.
This study refers to the former as the collapsing ``cloud'' and the latter as the star-forming ``core''.

If the mass of the high-density region exceeded the local Jeans mass, we judged them to be cloud/core.
We adopt the Bonner-Ebert mass \citep{Bonnor1956, Ebert1955}, $M_{\rm BE}$, as the local Jeans mass:
\begin{eqnarray}
    M_{\rm BE} &=& \frac{1.18 c^4_{\rm s}}{G^{3/2} P^{1/2}_{\rm ext}}\,\msun \,, \\ \label{eq:MBE}
    &\approx& 1050\,\msun\left(\frac{T}{200\,{\rm K}}\right)^{3/2} \left(\frac{\nh}{10^4\,\cc}\right)^{-1/2} \,, \label{eq:MBE2}
\end{eqnarray}
where $c_{\rm s}$ is the speed of sound, $G$ is the gravitational constant, $P_{\rm ext}$ is the external pressure, $T$ is the gas temperature, and $\nh$ is the hydrogen number density.
We calculate the local Jeans radius where $M_{\rm BE}(r) / M_{\rm enc}(r)$ is maximal, where $M_{\rm enc}(r)$ is the enclosed mass within a radius $r$ from the density centre of cloud/core.

We exclude the gravitationally trapped clouds/cores that move into the Jeans radius from the primary cloud/core because they eventually merge into the primary one.
We adopt $r_{\rm J} = 0.25$\,pc for the above classification, which is obtained by averaging the Jeans radius of the gas cloud obtained from these calculations.

\section{Numerical Simulations} \label{sec:results}

We study the number of first star-forming regions (clouds and cores) in $42$ cosmological simulations.
First, we summarise the effect of SV on physical properties at the virial scale (Section~\ref{sec:results:virial}) and the Jeans scale (Section~\ref{sec:results:jeans}), respectively.
Then we classify gas clouds into three types according to their morphology (Section~\ref{sec:results:type}) and show the number of clouds and cores (Section~\ref{sec:results:number}).
Table~\ref{t1} summarises the analysis results of $42$ models discussed in this section.

\subsection{Virial halo} \label{sec:results:virial}

\begin{figure*}
\begin{center}
\includegraphics[width=0.75\linewidth]{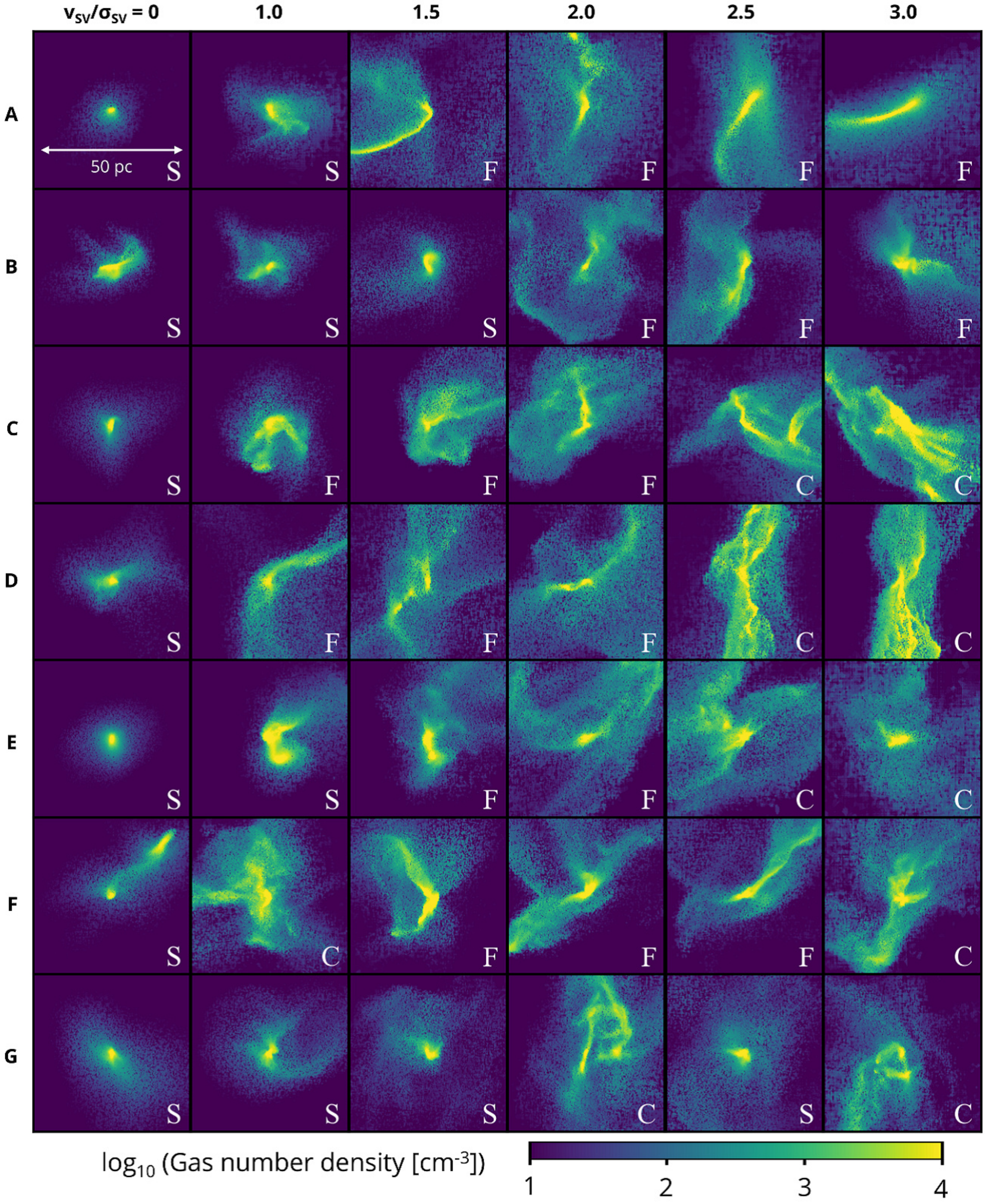}
\end{center}
\caption{
Distributions of the projected gas number density around the density peak for all models when the density first reaches $10^6\,\cc$.
The box sizes are $50\,$pc on a side.
Each model is placed on the parameter space of Halo A-G (vertical axis) and normalised initial SV $v_{\rm SV}/\sigma_{SV} = 0-3.0$ (horizontal axis).
The direction of the initial streaming velocity is aligned with the panel's horizontal axis (from left to right).
The letter at the bottom right of each panel indicates the classification of the gas cloud structure (Types~S, F, and C; see Section~\ref{sec:results:type}).
}
\label{fig:2Dmap_Density}
\end{figure*}

\begin{figure*}
\begin{center}
\includegraphics[width=0.75\linewidth]{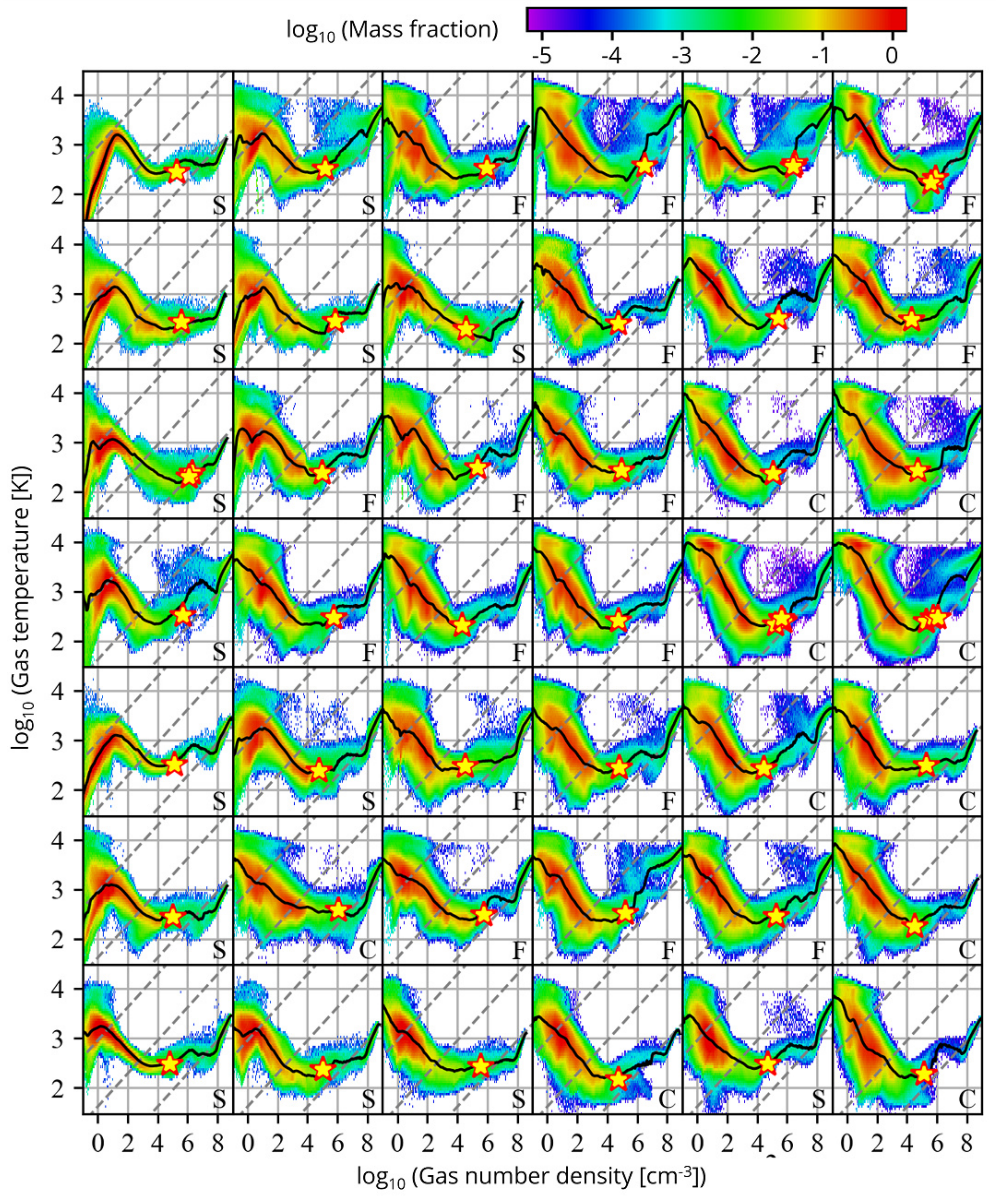}
\end{center}
\caption{
Phase diagrams of the gas temperature ($T$) as a function of the gas number density ($\nh$) for all models at the end of the simulations, $10^5$\,yr after the first core formation.
Each model is placed on the parameter space of Halo A-G (vertical axis) and normalised initial SV $v_{\rm SV}/\sigma_{SV} = 0-3.0$ (horizontal axis).
The colour map shows the distribution of the gas mass $\Delta M$ contained in the region of the logarithmic width $\Delta(\log \nh)$ and height $\Delta(\log T)$, where the redder the region, the larger the gas mass contained in it. 
The black lines show the logarithmic mean temperature weighted by the gas mass. 
The stars indicate the $\nh$-$T$ points where the collapsing cloud becomes gravitationally unstable.
The three dashed lines in each panel show the $\nh$-$T$ relation for the Jeans masses with $M_{\rm J} = 10^6$, $10^4$, and $10^2\,\msun$ (left to right), respectively.
The letter at the bottom right of each panel indicates the classification of the gas cloud structure (Types~S, F, and C; see Section~\ref{sec:results:type}).
}
\label{fig:42_rhoT}
\end{figure*}

\begin{figure*}
\begin{center}
\includegraphics[width=0.75\linewidth]{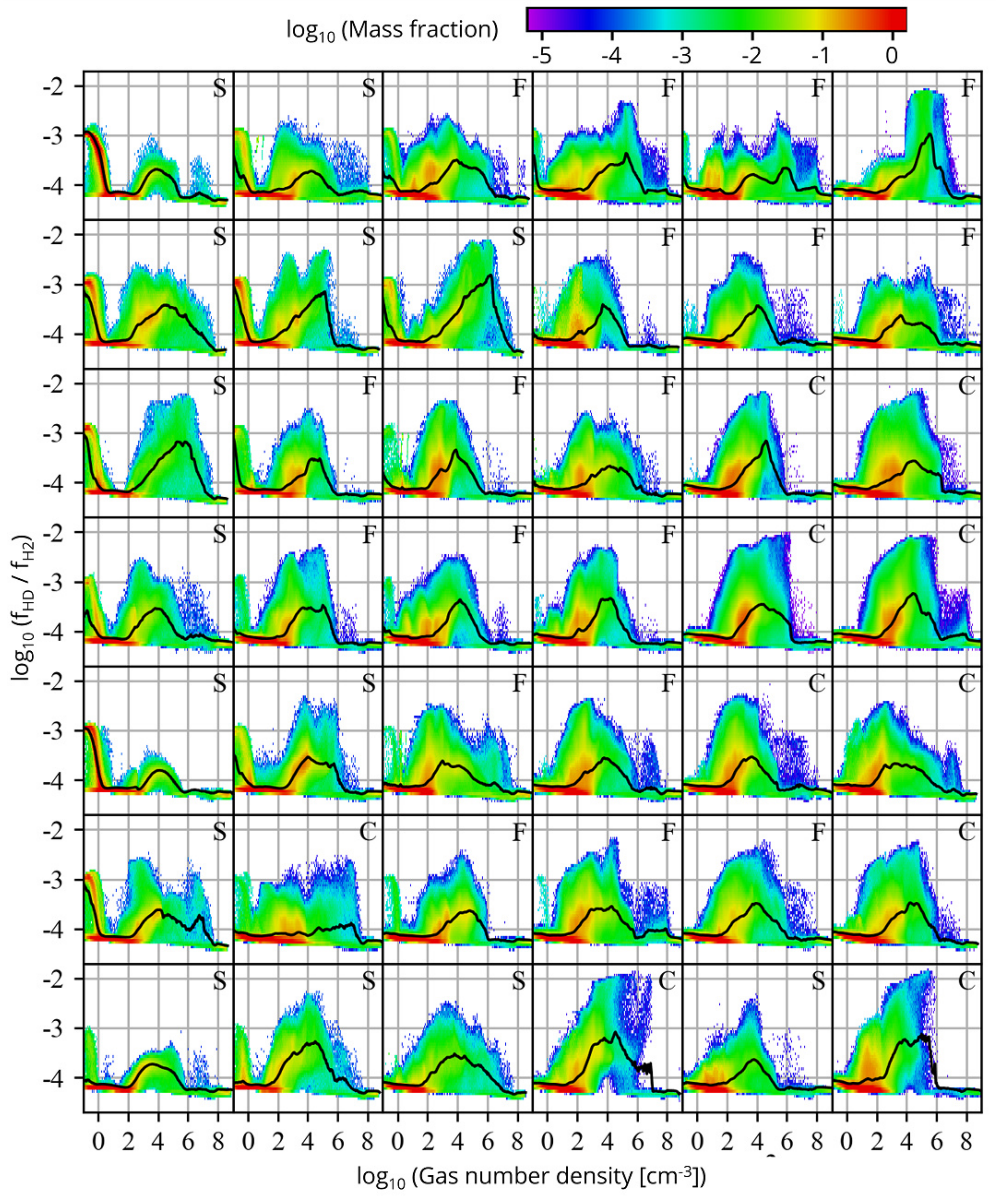}
\end{center}
\caption{
Same as Figure~\ref{fig:42_rhoT} but the vertical axis shows the abundance ratio $f_{\rm HD} / f_{\rm H_2}$.
}
\label{fig:42_HDH2}
\end{figure*}

Figure~\ref{fig:z-Mvir} shows the dependence of formation redshift and halo mass on the initial SV values.
As SV increases, star formation delays and halo mass increases (upper left direction).\footnote{Exceptionally, for models B25 and B30, this relationship is reversed with earlier formation epochs and smaller halo masses.}
This change is more significant for models that form at higher redshifts (i.e. in the order of Halo A to G).
Figures~\ref{fig:Vsv-Halo}(a) and \ref{fig:Vsv-Halo}(b) show the dependence of the redshift and halo mass on the model halos.
For example, the formation redshift is $dz = 13.37$ smaller, and the halo mass is $\Delta M_{\rm v} = 175$ times larger in Model~A30 compared with Model~A00, whereas $dz = 5.08$ and $\Delta M_{\rm v} = 8.5$ in Model~G30 compared with Model~G00.
As a result, the rates of $\Delta M_{\rm v}$ to $dz$ (the slope of lines in Figure~\ref{fig:z-Mvir}) are almost the same for A-G models, and on average, $\Delta M_{\rm v} \sim 100$ for $dz \sim 10$ between models with $v_{\rm SV} / \sigma_{\rm SV}^{\rm rec} = 0$ and $3.0$.

Besides delaying the star formation epoch, another effect of SV on the halo scale phenomena is to change the gas mass fraction, the baryon fraction $f_{\rm b} = M_{\rm v,b} / M_{\rm v}$.
Figure~\ref{fig:Vsv-Halo}(c) shows the dependence of the baryon fraction on the initial SV.
The baryon fraction becomes smaller for halos born at higher redshift.
The higher redshift models, e.g. Halos~A and B, have lower baryon fractions regardless of SV, while the lower redshift models, e.g. Halo~G, have higher baryon fractions.
In the intermediate models, Halos~C to F, the baryon fractions tend to decrease with SV (negative correlation).
Figure~\ref{fig:Vsvcol-fb} shows the dependence on SV at the collapse time (redshift in Table~\ref{t1}). The baryon fraction decreases with SV where SV at the collapse time is above about 0.03, except in some models.
We attribute the increase in baryon fractions with SV in some models to the increase in the gravitational potential of the halo, which allows the inflow gas to remain in the halo without leaking out.

Figure~\ref{fig:2Dmap_Density} shows the gas density distribution inside each halo.
In the absence of SV (left-most column), the gas in the halo contracts while maintaining a spherically symmetric structure.
As SV increases, the gas structure deviates from spherical symmetry, and an elongated filamentary or sheet-like structure appears.
There are two possible mechanisms by which SV changes the density structure of the gas inside the DM halo.
\begin{itemize}
    \item[I.] Crushing orthogonal to the initial SV direction by the gravitationally bound gas flowing into the halo.
    \item[II.] Spreading parallel to the initial SV direction by gas that is not gravitationally bound and flows out of the halo.
\end{itemize}
Two model parameters change the degree of the above two effects: the larger $M_{\rm v}$, the more substantial effect I appears, while the faster SV, the more substantial effect II appears.
As a result, various gas density structures appear.

\begin{figure*}
\begin{center}
\includegraphics[width=0.9\linewidth]{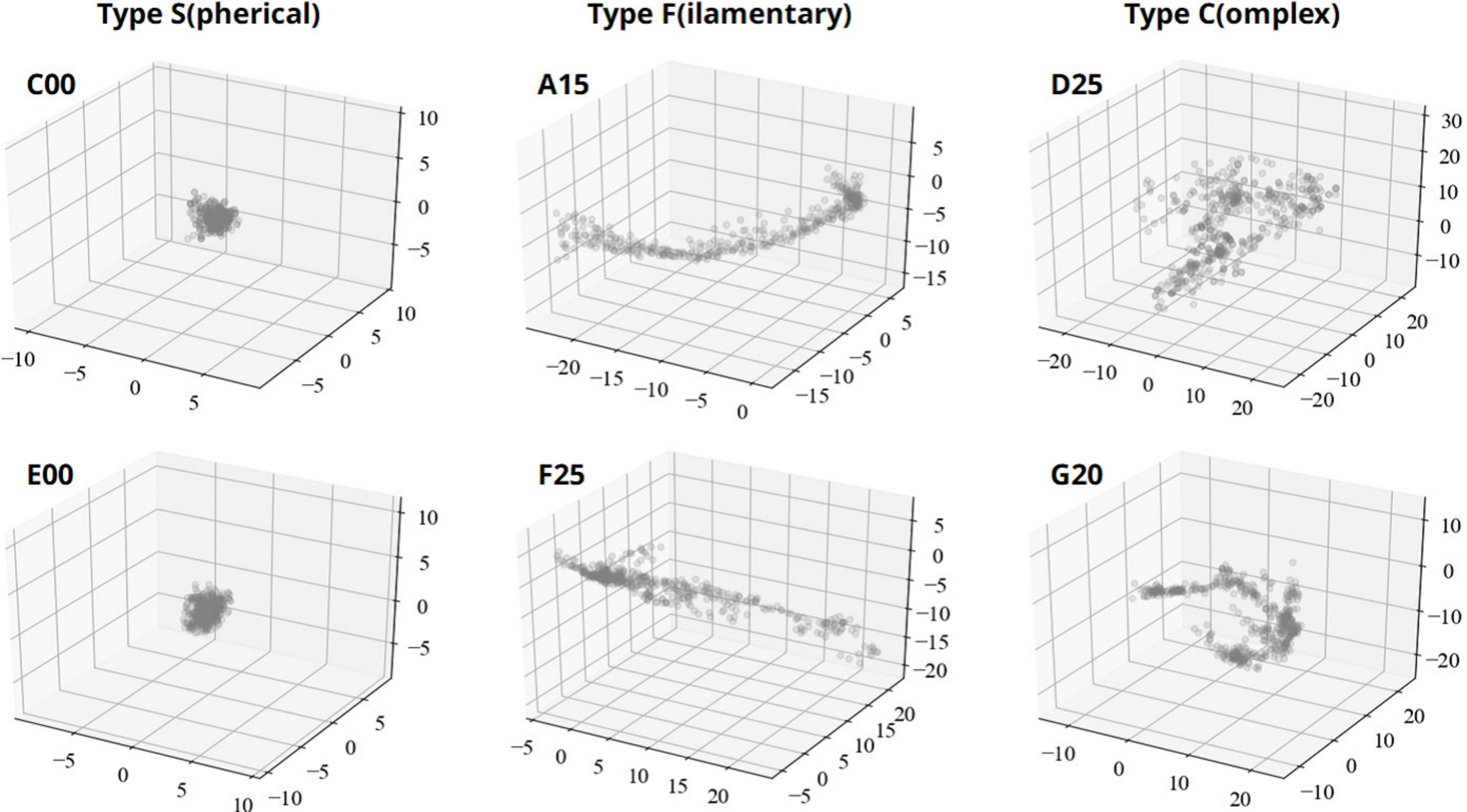}
\end{center}
\caption{
3D Distributions of the gas (SPH) particles with $\nh \simeq 10^3\,\cc$, which are used to classify the cloud's morphology.
The left panels show Type~S (spherical), Models~C00 and E00.
The middle panels show Type~F (filamentary), Models~A15 and F25.
The right panels show Type~C (complex), Models~D25 and G20.
The unit of the axis in all panels is parsec.
}
\label{fig:Type}
\end{figure*}

\begin{figure}
\begin{center}
\includegraphics[width=0.9\columnwidth]{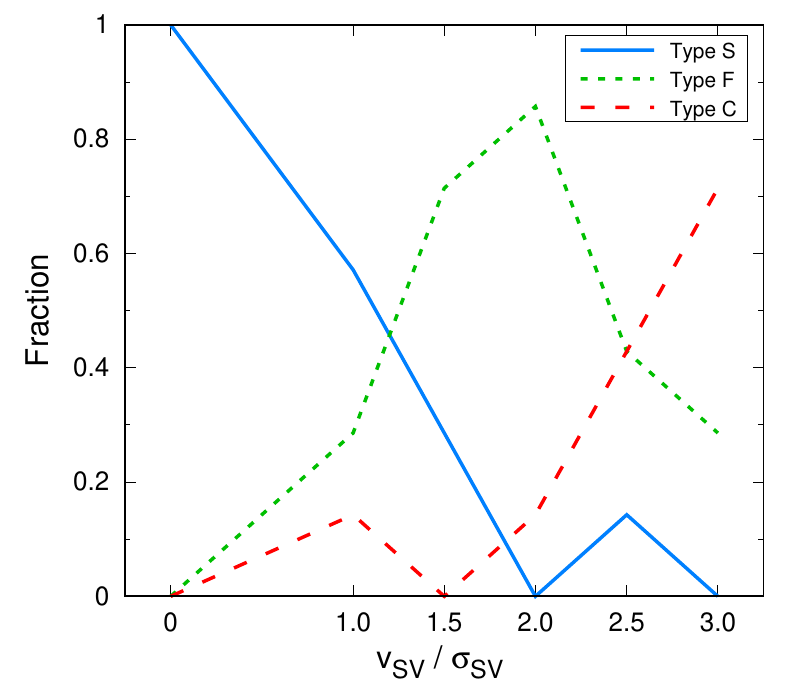}
\end{center}
\caption{
Fraction of the cloud's structure classes (Types~S, F, and C) as a function of SV.
The solid, dotted, and dashed lines correspond to Type~S, F, and C, respectively.
}
\label{fig:bars}
\end{figure}

\begin{figure}
\begin{center}
\includegraphics[width=0.9\columnwidth]{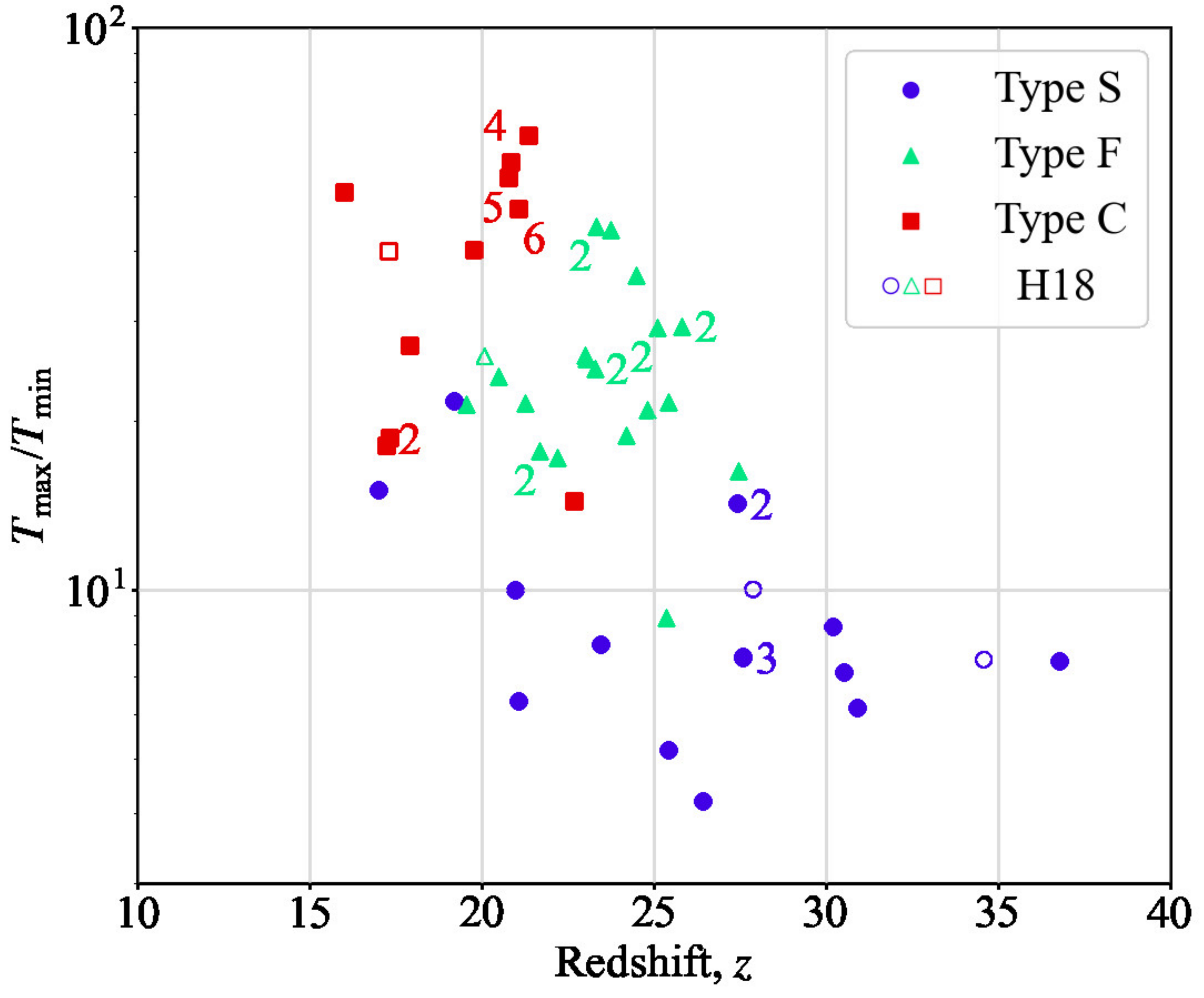}
\end{center}
\caption{
The ratio of the maximum and minimum temperatures of the collapsing gas cloud, $T_{\rm max}/T_{\rm min}$, as a function of the redshift.
The temperatures are calculated from the average track (black solid lines in Figure~\ref{fig:42_rhoT}) with $\nh = 1 - 10^6\,\cc$.
The circles, triangles, and squares correspond to Type~S, F, and C, respectively.
The filled symbols are for the models in this study.
The open symbols are for the models in \citetalias{Hirano2018}, for comparison.
Models in which multiple clouds ($\nh \geq 10^6\,\cc$) are detected have the number of clumps beside the symbol.
}
\label{fig:sigma-zTratio}
\end{figure}

\subsection{Jeans cloud} \label{sec:results:jeans}

We set $\tth = 0$ when the gas number density at the centre of the collapsing cloud first reaches $\nh = \nth = 10^8\,\cc$ and continue the simulation until $\tth = 10^5$\,yr using the opaque core methodology (Section~\ref{sec:method:hydrosim}).

Figure~\ref{fig:42_rhoT} displays the phase diagrams on the density-temperature ($\nh-T$) plane of all models at the end of the simulation ($\tth = 10^5$\,yr).
On the lower density region ($\nh \lesssim 10\,\cc$), the gas fluid is contracted by the gravity of the DM halo and is adiabatically compressed.
Since the DM halo mass increases with SV, both the mass and temperature of the gas material inside the DM halo increase with SV (from left to right panels in Figure~\ref{fig:42_rhoT}).

After the molecular hydrogen (H$_2$) forms, H$_2$ radiative cooling exceeds the compression heating, and the gas temperature decreases ($10 \lesssim \nh/\cc \lesssim 10^5$).
As a result of the temperature drop, the pressure also decreases, which leads to a self-gravitational contraction of the gas cloud.

The thermal evolution of the collapsing gas cloud goes from temperature decreasing to increasing around $\nh \sim 10^5\,\cc$
At the point at which the gas temperature reaches a minimum (``loitering point''), the gas cloud becomes a gravitationally unstable state (see star symbols in Figure~\ref{fig:42_rhoT}).
After that, the gas cloud can gravitationally contract while increasing gas temperature.

Besides H$_2$-cooling, hydrogen deuteride (HD)-cooling is also vital for metal-free star formation.
Previous studies suggested that HD-cooling becomes influential on the thermal evolution of the collapsing cloud if the abundance ratio $f_{\rm HD} / f_{\rm H_2}$ overcomes $10^{-3}$ \citep{Ripamonti2007}.
Figure~\ref{fig:42_HDH2} shows the phase diagrams of the ratio for all models.
The abundance ratio exceeds the threshold value in some models (e.g. A30, B15, D30).
The temperature in those models sharply decreases at $\nh = 10^4 - 10^6\,\cc$ (Figure~\ref{fig:42_rhoT}), which can affect the fragmentation scale during the long-term evolution.

The above phenomena occur in a density lower than the threshold value $\nth = 10^8\,\cc$. 
When the gas density exceeds the threshold density, the temperature artificially increases due to the opaque core methodology.

\subsection{Structure classification} \label{sec:results:type}

Figure~\ref{fig:2Dmap_Density} displays various 3D morphologies of the collapsing gas clouds depending on the initial SV values.
The shape of the gas cloud, particularly the degree of elongation, relates to the (filament) fragmentation process.
We classify the shape of gas clouds into the following three types, Types~S, F, and C.
We distinguish them using the ratio of the major and minor axes ($a/b$) of the iso-density surface at $\nh = 10^3\,\cc$.
Figure \ref{fig:Type} shows examples of the three distinct shapes.
\begin{itemize}
    \item[S] (spherical): iso-density surfaces with $a/b < 3$, which includes structures from spheres to ellipsoids. We henceforth classify them as spherical structures to distinguish them from the filamentary structure discussed below.
    \item[F] (filamentary): iso-density surfaces are approximated as an elongated cylinder with $a/b > 3$. 
    \item[C] (complex): iso-density surfaces consist of multiple filaments and/or sheets and are approximated as neither sphere nor cylinder.
\end{itemize}

Figure~\ref{fig:bars} summarises the dependence of the cloud structure on SV.
The cloud structure is strongly influenced by the initial SV of the forming region: Types~S, F, and C are dominant at low ($v_{\rm SV}/\sigma_{SV}\leq1.0$), intermediate ($1.0 \leq v_{\rm SV}/\sigma_{SV}\leq2.5$), and high ($v_{\rm SV}/\sigma_{SV}\geq2.5$) SV models, respectively.
The clouds of Type~S have the same structure, and the iso-density surface sizes are within the radius of $5$\,pc.
In contrast, for the clouds of Type~F, which have elongated structures, the thickness of the filaments is about $3-5$\,pc and the length is distributed around $20-40$\,pc.
Some of the filaments had large curvatures or were split in the middle.
For the clouds of Type~C, we find multiple filaments intertwined within a structure extending over several tens of pc.

When the cloud contracts while lowering its temperature, the asymmetry of the cloud structure increases due to progressive contraction in a specific direction (dimension).
We focus on the magnitude of the temperature drop of the collapsing cloud to explain the three types of cloud morphologies.
The maximum temperature at the virial scale depends on the halo mass and increases with SV ($\sim\!10^3 - 10^4$\,K).
The minimum temperature at the Jeans scale (the ``loitering point'') depends on the coolants; H$_2$ can cool down to $200$\,K and HD to $50$\,K.
Therefore, the temperature drops generally increase as SV increases (Figure~\ref{fig:42_rhoT}).

Figure~\ref{fig:sigma-zTratio} shows the ratio of the maximum and minimum temperatures on the average track (solid black lines in Figure~\ref{fig:42_rhoT}) during $\nh = 1 - 10^6\,\cc$.
The asymmetry/complexity of gas clouds tends to increase (Type S $\to$ F $\to$ C) as the temperature ratio increases.
The temperature ratio $T_{\rm max}/T_{\rm min} = 10 - 15$ is a threshold, below which the cloud is symmetry (Type S) whereas above which asymmetry (Types F and C).
Above the threshold ratio, the formation redshift becomes an additional parameter: the lower the redshift, we classify the models with relatively small temperature ratios as Type~C.\footnote{Model~F10 ($z = 22.70$ and $T_{\rm max}/T_{\rm min} = 15$) is classified as Type~C despite having a small SV and a relatively small temperature ratio. It has a short side chain branching from the centre of the filament, but the actual structure is similar to Type~F.}

\subsection{Numbers of clouds and cores} \label{sec:results:number}

\begin{figure*}
\begin{center}
\includegraphics[width=0.8\linewidth]{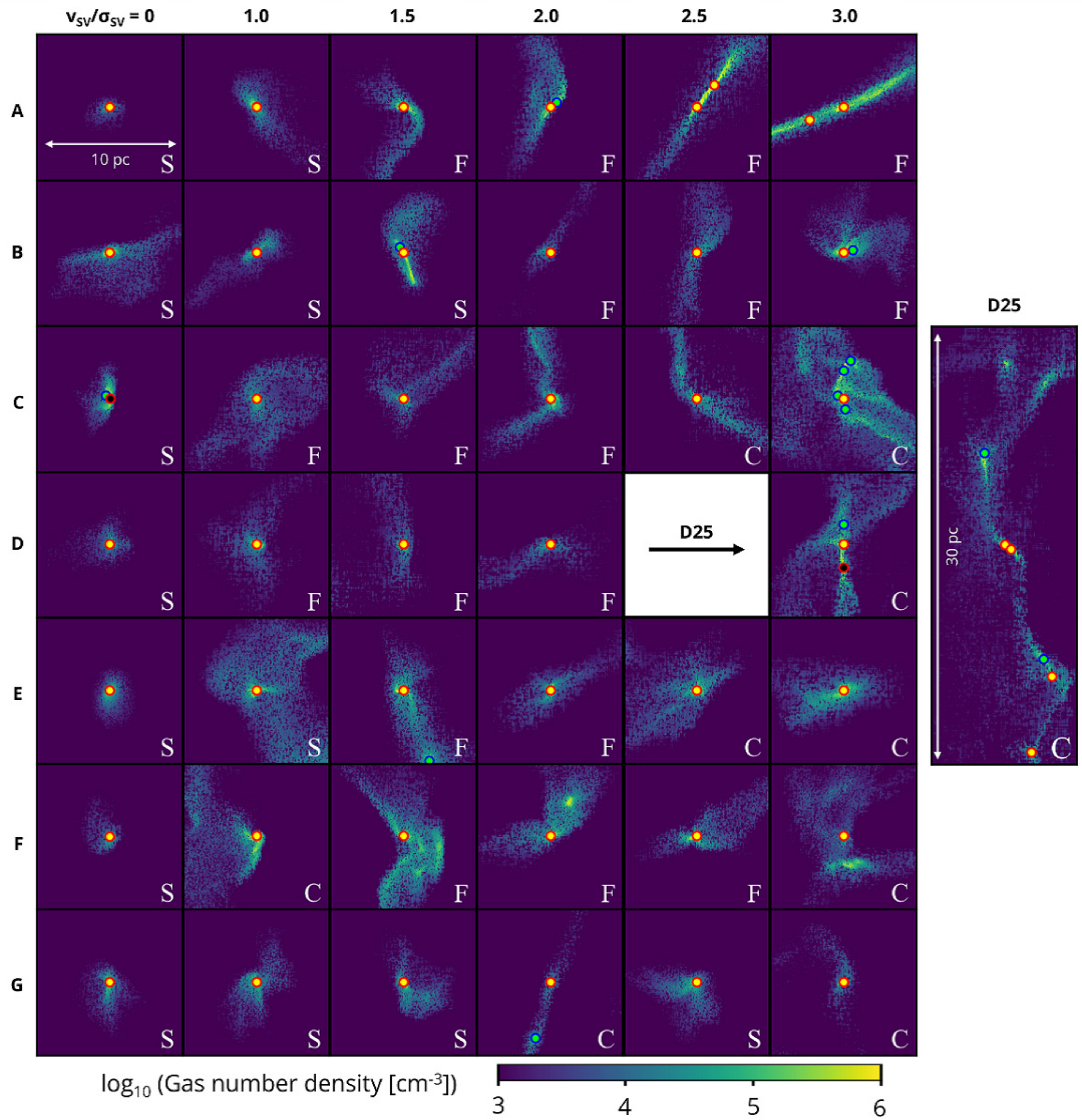}
\end{center}
\caption{
Distributions of the projected gas number density around the density peak at the end of the simulations ($\tth = 10^5$\,yr).
We plot each model on the parameter space of Halos~A-G (vertical axis) and normalised initial SV $v_{\rm SV}/\sigma_{SV} = 0-3.0$ (horizontal axis).
The box sizes are $10\,$pc on a side.
The red circles show cores (with $\nh \geq  10^8\,\cc$) and mean that a core forms by contracting inside the cloud.
The blue circles show clouds (with $\nh \geq  10^6\,\cc$) and mean that a cloud forms, but a core has not yet formed inside the cloud.
We blacked out the inside of the circle to indicate close pairs of cores ($r<r_{\rm J} = 0.25$\,pc) found in Models~C00 and D30.
Model~D25 has an elongated cloud then we show the entire structure in the widened panel ($30$\,pc along the vertical axis) outside the right side of the figure.
The letter at the bottom right of each panel indicates the classification of the cloud structure (Types~S, F, and C).
}
\label{f11}
\end{figure*}

To identify the formation sites of the first stars, we adopt two physical conditions: (1) ``clouds'' with $\nh \geq 10^6\,\cc$ where the collapsing cloud becomes gravitationally unstable and (2) ``cores'' with $\nh \geq10^8\,\cc$ which is the threshold density of the opaque core technique.
Figure~\ref{f11} shows distributions of the clouds (blue circles) and cores (red circles) at the end of the simulations.
Multiple clouds and cores formations occur, especially on elongated filamentary structures, even in Type~C models.

\begin{figure}
\begin{center}
\begin{tabular}{cc}
\includegraphics[width=0.9\linewidth]{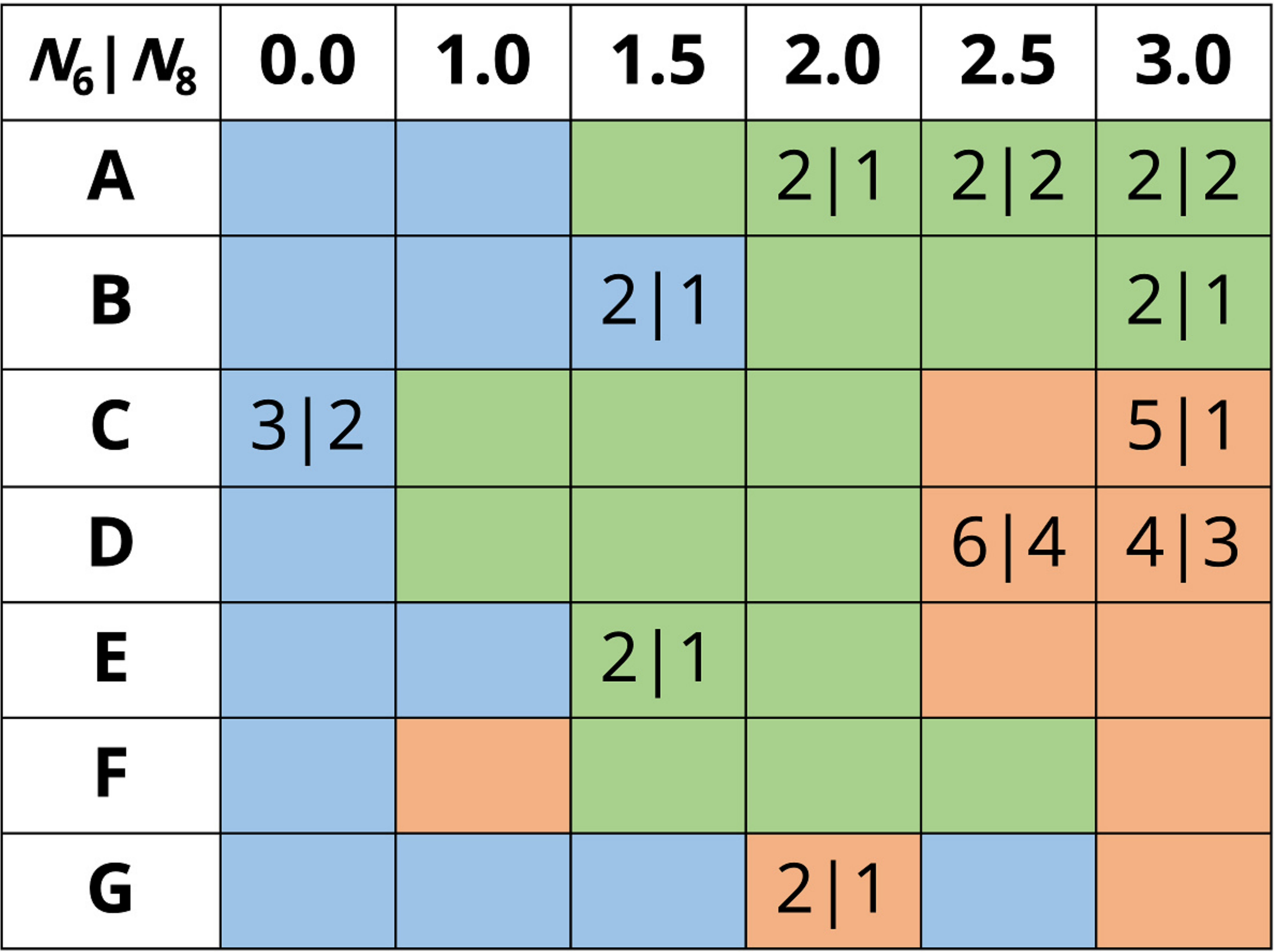}
\end{tabular}
\end{center}
\caption{
Numbers of clouds ($N_6$ where $\nh \geq 10^6\,\cc$) and cores ($N_8$ where $\nh \geq 10^8\,\cc$) at the end of the simulations ($\tth = 10^5$\,yr).
We omit the number ``1'' and leave the cell blank.
The cell colours indicate the structure class of the clouds: blue, green, and red correspond to Types~S, F, and C, respectively.
}
\label{f12}
\end{figure}

Figure~\ref{f12} summarises the number of clouds and cores.
The colours of cells distinguish the classes of cloud structure.
The more complex the cloud structure is (Type~S $\to$ F $\to$ C), the larger the number of clouds and cores. 
\begin{itemize}
    \item[S]
    (spherical): Most of the gas clouds in this class form a single core at the centre of the spherically symmetric collapsing gas clouds.
    This is consistent with the typical scenario in the first star formation.
    The exceptions are Models~B15 and C00.
    Model~B15, in which two clouds and one core are detected, is considered an intermediate state between Types~S and F.
    The gas cloud of Model~B15 has a higher oblateness than other Type~S models and fragments into two clouds. 
    Model~C00, in which three clouds and two cores are detected, on the other hand, is exceptional. 
    The multiple clouds and cores form inside the spherically symmetric collapsing gas cloud despite the absence of SV.
    Because two cores are born near the density centre, they approach a distance less than the local Jeans length ($r_{\rm J} = 0.25$\,pc is the average value for the collapsing gas clouds in this study) at $\tth = 0.8 \times 10^5$\,yr.
    We distinguish them as a close pair of cores as the black circles in Figure~\ref{f11}.
    \item[F]
    (filamentary): A maximum of two objects are formed along the filamentary gas cloud.
    Models~A20, B30, and E15 form a pair of cloud and core with distances $0.66$, $0.63$, and $5.2$\,pc, respectively.
    Models~A25 and A30 form a pair of two cores with distances $1.9$ and $2.6$\,pc, respectively.
    The fragmentation scale corresponds to the Jeans length $r_{\rm J}=2$\,pc at $\nh = 10^4\,\cc$.
    \item[C]
    (complex): A maximum of six objects are formed in the complex gas cloud.
    In these cases, clouds and cores form in a row on a dense filament, ranging from about $5$\,pc to more than $20$\,pc in length.
    Model~D25 has the most significant number of clouds (six) and cores (four) examined in this paper.
    Model~D30 has a close pair of cores formed by approaching within the Jeans length ($r_{\rm J} = 0.25$\,pc) at $\tth = 0.9 \times 10^5$\,yr
\end{itemize}

The analytical study of the filament fragmentation \citep{Inutsuka1992, Inutsuka1997} showed that the filament thickness is approximately equal to the Jeans length and the filament fragments at every Jeans length.
The models classified as Types~F and C in this study have dense filaments.
However, the number of clouds and cores is smaller than that for the case of splitting at each Jeans length.
In most models, dense filaments form only one core without fragmentation.

\section{Discussion} \label{sec:dis}

\subsection{Star formation efficiency} \label{sec:dis:Mstar}

\begin{figure*}
\begin{center}
\includegraphics[width=0.9\linewidth]{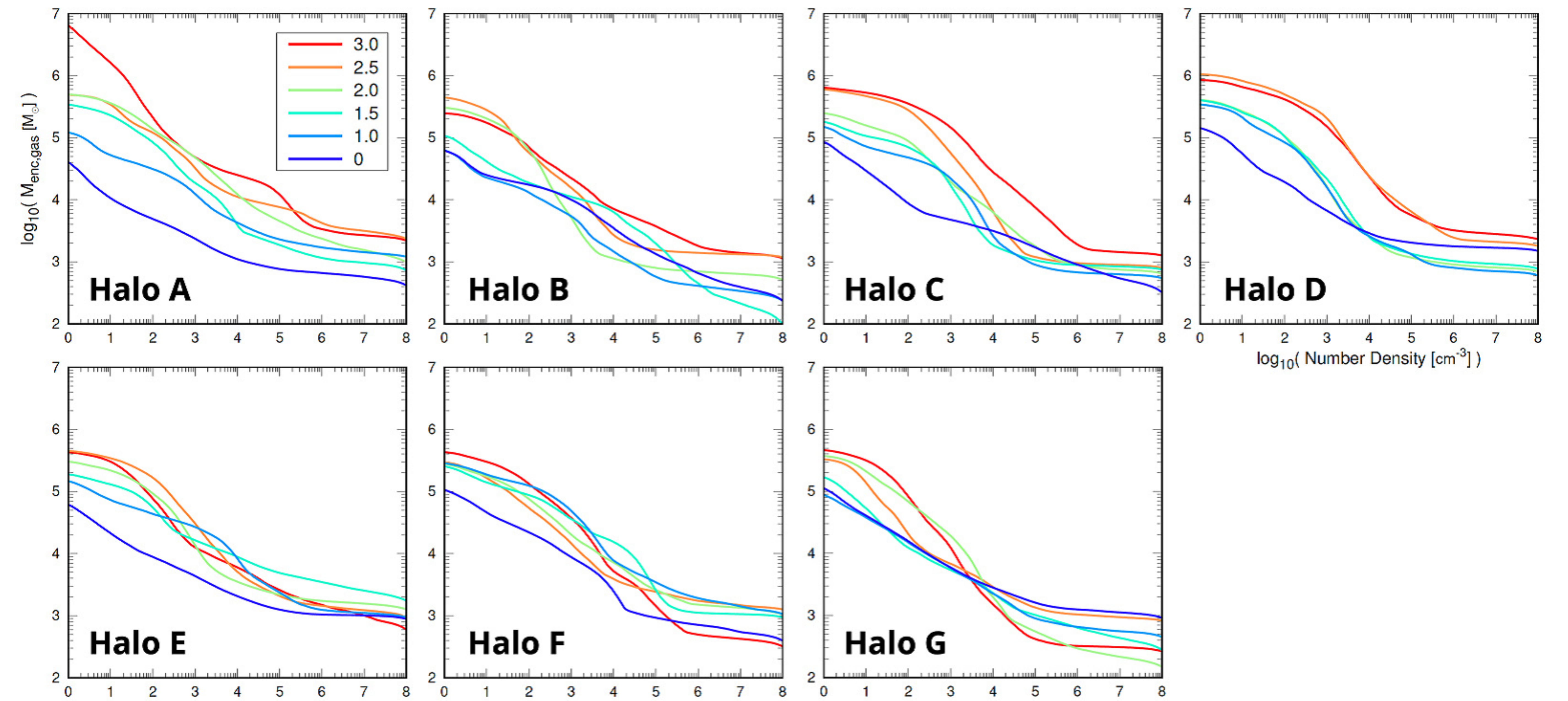}
\end{center}
\caption{
Distributions of the enclosed gas mass as a function of the gas number density, $M_{\rm enc,gas}(>\!\nh)$, at the end of the simulation ($\tth = 10^5$\,yr).
Each panel shows Halos~A to G results with different initial SV $v_{\rm SV}/\sigma_{\rm SV} = 0-3.0$.
}
\label{fig:Dens-Menc}
\end{figure*}

\begin{figure}
\begin{center}
\includegraphics[width=0.9\columnwidth]{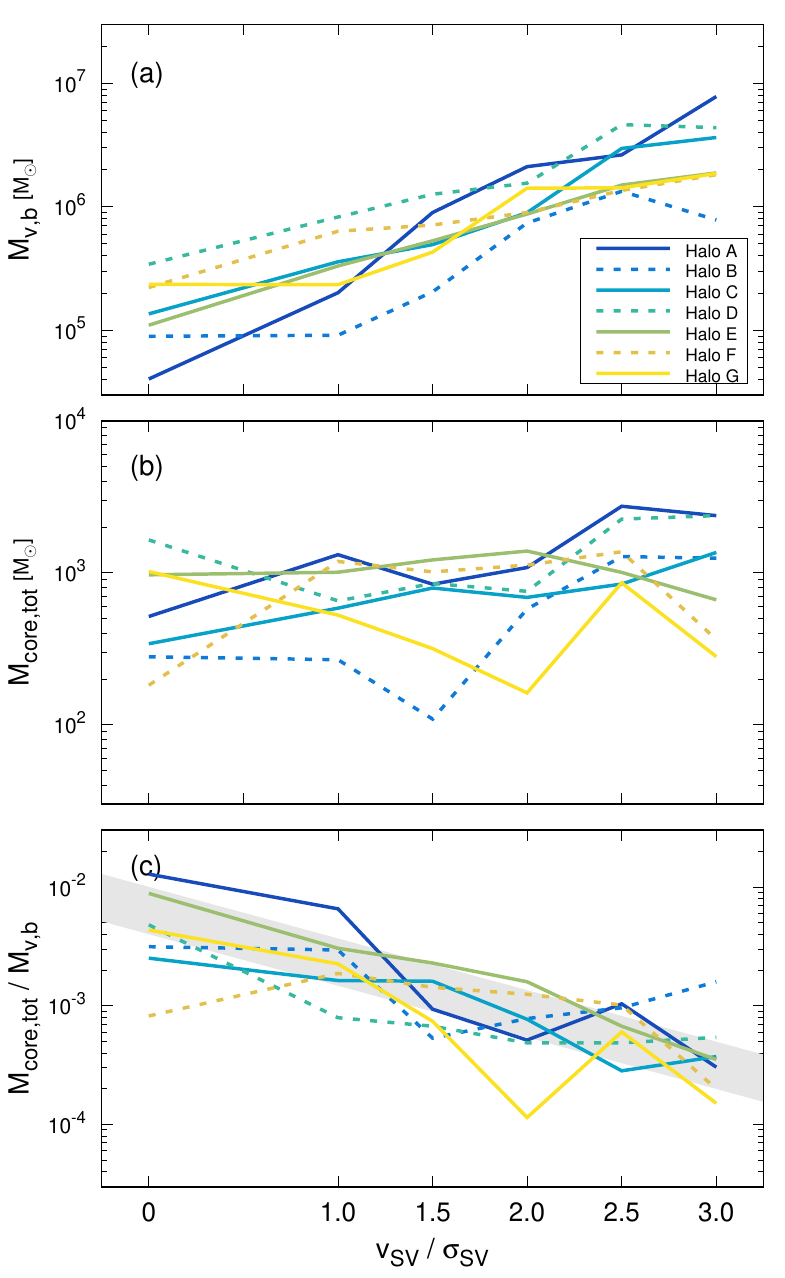}
\end{center}
\caption{
Gas mass as a possible mass fuel for the first star formation as a function of the initial SV.
Panels: (a) gas mass inside the virial radius, (b) total core masses where $\nh \geq 10^8\,\cc$, and (c) ratio of the above two masses.
The grey-shaded region in panel (c) corresponds to Equation~\ref{eq:EscapeFraction}.
}
\label{fig:Vsv-Cloud}
\end{figure}

To calculate the long-time thermodynamic evolution of the massive gas cloud of $10^4-10^5\,\msun$, we limit the maximum computational density to $\nth = 10^8\,\cc$.
Since this is lower than the formation density of protostars ($\sim\!10^{20}\,\cc$), the current simulations can not indicate the masses of primary stars born inside individual cores.

On the other hand, since the first stars form at the centre of the identified cores, the core mass constrains an upper limit of the first star mass.
If we consider that the core mass correlates with the masses of the first stars, it makes sense to examine the dependence of the core mass on the model parameters (host halo and initial SV).

Figure~\ref{fig:Dens-Menc} shows the enclosed gas mass for each model as a function of the gas number density.
On the low-density side (large-scale), which corresponds to the early stages of star formation, the enclosed gas mass increases with increasing initial SV for all models.
The delay of the gas contraction (beginning of the star formation) in the halo due to SV causes this correlation and increases gas mass at the virial scale ($M_{\rm v,b}$ in Figure~\ref{fig:Vsv-Cloud}(a)).

However, as the gas contraction proceeds, the SV dependence of the gas mass is more varied on the high-density side (small-scale) in Figure~\ref{fig:Dens-Menc}.
For models related to Halo~A, the gas clouds contract with maintaining the positive SV correlation.
For Halos~B-D, models initiated with high SV also show a positive correlation.
On the other hand, gas masses for models initiated with intermediate SV go below that for the model without SV on the high-density side.
Halos~E-G show that the SV dependence of the gas mass reverses from positive to negative during contraction.

At the scale where the contracting gas cloud first becomes gravitationally unstable as with sphere, filament, and sheet-like structures, the gas mass of such objects ($M_{\rm J,whole}$ in Table~\ref{t1}) shows a monotonic increase with SV.
For larger SV, $M_{\rm J,whole}$ increases, and $M_{\rm J,whole}$ increases about $40$ times on average for $v_{\rm sv}/\sigma_{\rm sv} = 0\to3.0$.
Gravity contraction progresses, and clouds/cores form inside these objects.
We can confirm that local Jeans mass around each core ($M_{\rm J}$) does not always correlate to the Jeans mass of the parent cloud ($M_{\rm J,whole}$).

Figure~\ref{fig:Vsv-Cloud}(b) summarises the total core mass, summation of core masses for each model, at the end of simulations.
On average, the dependence on SV for total core mass is weak.
The weakening dependence on SV means that the ratio of the total core mass to the virial gas mass decreases with SV (Figure~\ref{fig:Vsv-Cloud}(c)).
Let us consider this ratio to represent the upper limit of star formation efficiency used in the context of galaxy formation.
The formation efficiency decreases with SV by more than one order of magnitude (see the grey-shaded region in Figure~\ref{fig:Vsv-Cloud}(c)),
\begin{equation}
  \epsilon_{\rm III} \propto \frac{M_{\rm core,tot}}{M_{\rm v,b}} = (0.004 \sim 0.01) \cdot \exp \left( -\frac{v_{\rm sv}}{\sigma_{\rm sv}} \right)\,.
  \label{eq:EscapeFraction}
\end{equation}
Massive gas clouds born under large SVs do not necessarily form many first stars.

\subsection{Supermassive star} \label{sec:dis:SMBH}

Among the models in this study, the maximum mass of the dense core is $M_{\rm core} \sim 1400\,\msun$, and the total value for each model is $M_{\rm core,tot} \sim 3000\,\msun$ at most.
The studied models have no supermassive stars with $\sim\!10^5\,\msun$ that collapse directly into the intermediate-mass BHs (IMBHs).

There are several models with $M_{\rm J,whole} = 10^4-10^5\,\msun$ (Table~\ref{t1}).
If a protostar efficiently acquires mass in such a gas cloud, the protostellar radiative feedback, which can set the final stellar mass by terminating the mass accretion, will not work \citep{McKee2008, Hosokawa2011}.
In this case, the protostar may grow to a large mass over $1$ million years.
However, the present calculation only estimates the core mass at $10^5$\,yr after the formation of the high-density region (which corresponds to the protostar formation epoch), and these simulations do not suppress the upper limit of the final stellar mass in a high-accretion environment.

We analysed the radially averaged accretion rate centred on the dense core to obtain the total gas mass in the range above the critical mass accretion rate required to suppress the protostellar radiative feedback, $\dot{M}_{\rm cr} = 0.047\,\msunyr$ \citep{Hosokawa2012a}.
Since there is no mechanism to suppress gas accretion without radiation feedback while this gas is accreting to the protostar, we can consider this gas mass as the stellar mass, assuming that the protostar continues to grow in mass.
Among the masses estimated in this way, Model~B30 has the most enormous value, $9233\,\msun$, similar to the intermediate massive first stars found in the previous studies \citep[e.g.,][]{Hirano2017a, Wise2019, Regan2020}.
Some of the models investigated in this study may form IMBHs, which will be clarified by examining the results of long-time calculations to be performed in the future.

\cite{Hirano2017a}, directly calculated the formation of a supermassive first star, selected a target DM halo whose central velocity dispersion is $\sim\!160\,\kms$ at $z = 7$, which is consistent with the estimated value of the host galaxies of observed high-$z$ SMBHs.
In the target halo initiated with $v_{\rm SV}/\sigma_{\rm SV}=3.0$, a massive first star with $3.4\times10^4\,\msun$ was formed at $z = 30.5$ inside a massive DM halo with $M_{\rm v}=2.2\times10^7\,\msun$ (a star symbol in Figure~\ref{fig:z-Mvir}).
If we find samples of supermassive first stars forming in the models we have investigated in this paper, we can relax the formation conditions.
Further systematic studies calculating longer-term evolution are needed to clarify to what extent the conditions for forming supermassive first stars are acceptable.

\subsection{Multiple and binary} \label{sec:dis:binary}

Another issue of this series of studies is whether massive gas clouds formed by delayed star formation due to SV become gravitationally unstable and fragment.
Five models in this study show the formation of multiple cores ($N_8>1$; Table~\ref{t1}).
Four cores formed in Model~D25 ($N_8 = 4$), the maximum number of multiple cores for each model in this study, and their masses vary from $52$ to $792\,\msun$ each.
The four cores exist over a long filament (Figure~\ref{f11}).
The cores (red circles) and clouds (blue circles, overplotted by red ones) are all found on filamentary structures.

Two of the models with multiple core formations identified two cores approaching within the Jeans length ($r_{\rm J} = 0.25$\,pc) at the end of the calculation (asterisks in Table~\ref{t1} and black circles in Figure~\ref{f11}).
\begin{itemize}
    \item Model~C00: $M_{\rm core} = 234$ and $107\,\msun$
    \item Model~D30: $M_{\rm core} = 759$ and $524\,\msun$
\end{itemize}
In Model~D30, given the large SV, the core mass, which is the upper limit of stellar mass, is also increased.

Since these simulations do not numerically resolve the individual stellar scales, whether these will eventually build a close binary pair is under debate.
Also, due to the limitations of this study's numerical resolution and output time increments, we may have missed clouds/cores that merged in the process.
To uncover the formation and evolution of multiple clouds/cores, high-resolution and long-term simulations are necessary for the future.

\subsection{Numerical resolution} \label{sec:dis:resolution}

Finally, we discuss the dependence of the results on the numerical resolution.
In this study, we limit the maximum density to the threshold density $\nth$, whose corresponding scale is smaller than $0.1$\,pc, to perform the calculation of the first $10^5$\,yr evolution after during the accretion phase.
Whether the number of fragments increases when calculating up to higher densities (smaller scales) is a factor that determines the applicability of the results of this study.

\begin{figure*}
\begin{center}
\includegraphics[width=0.9\linewidth]{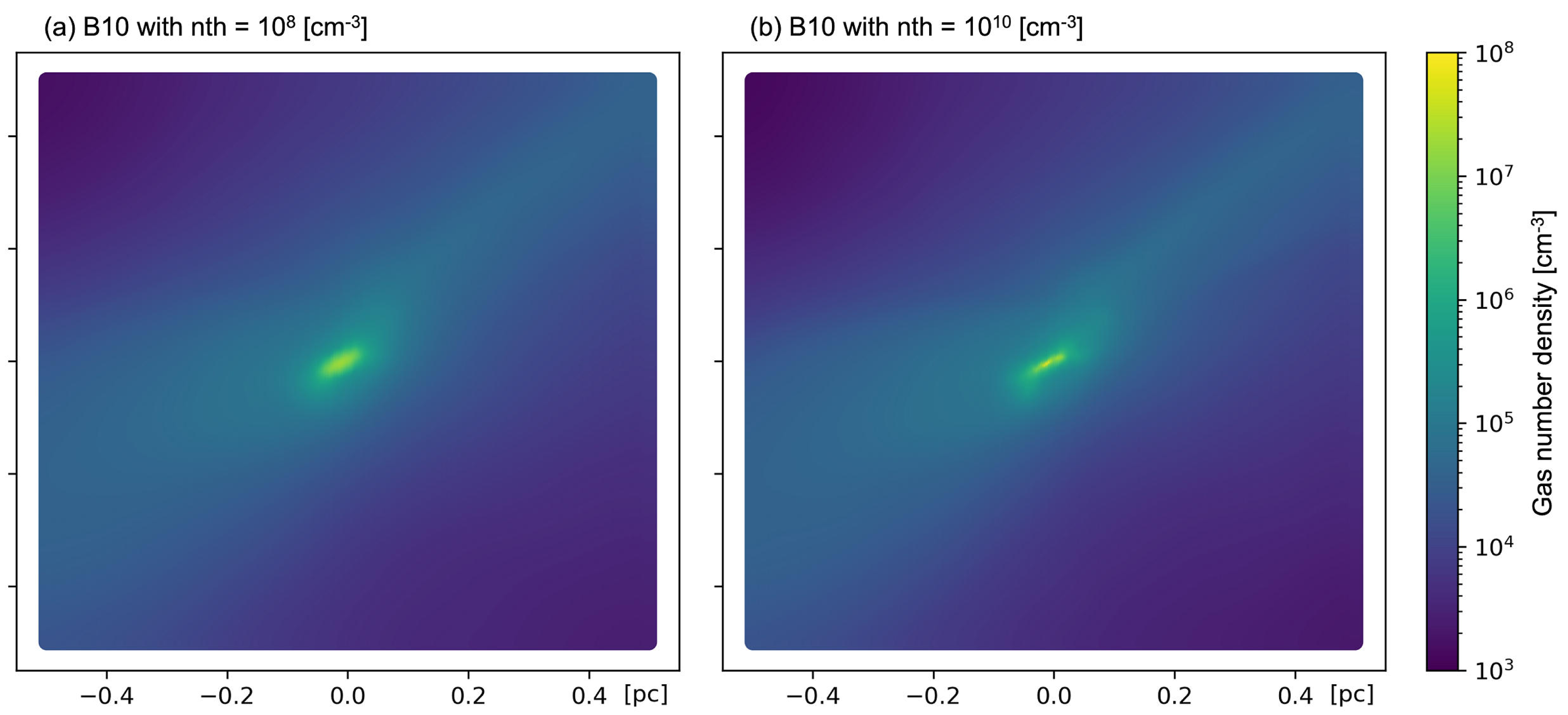}
\end{center}
\caption{
Distributions of the projected gas number density around the density peak at the end of the simulations ($\tth = 10^5$\,yr): (left) Model~B10 with $\nth = 10^8\,\cc$ and (right) Model~B10 with $\nth = 10^{10}\,\cc$.
The box sizes are $1\,$pc on a side.
}
\label{fig:14}
\end{figure*}

\begin{figure}
\begin{center}
\includegraphics[width=0.9\linewidth]{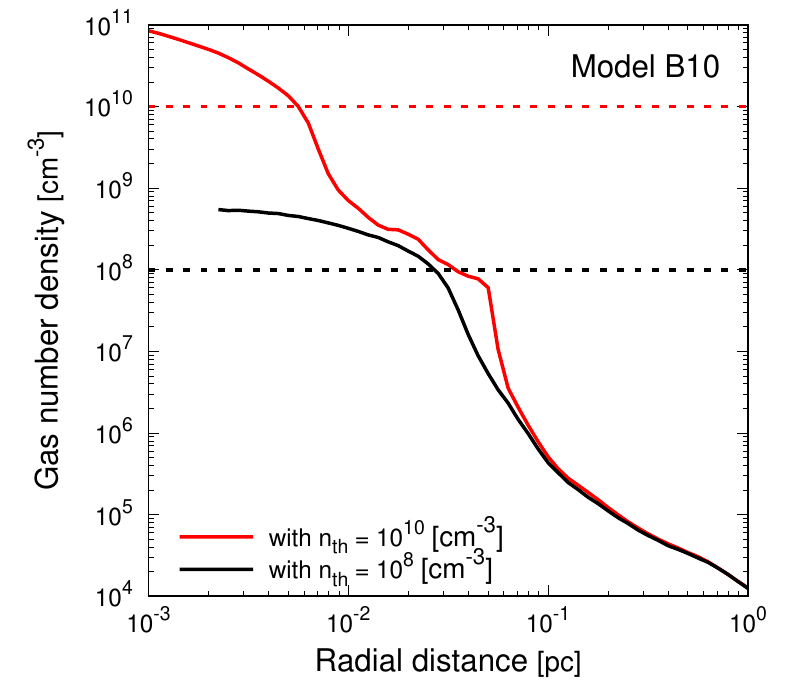}
\end{center}
\caption{
Radial gas number density profiles for Model~B10 with different threshold densities, $\nth = 10^8$ and $10^{10}\,\cc$ at the end of the simulation ($\tth = 10^5$\,yr).
The horizontal dashed lines show the threshold densities for each run.
}
\label{fig:15}
\end{figure}

To confirm the dependence on numerical resolution, we simulate Model~B10 with $\nth =10^{10}\,\cc$.
Figure~\ref{fig:14} compares the gas density distributions for simulations with different $\nth$, and there is no apparent difference in the overall structure.
Since the higher computational resolution allowed for smaller structures to be constructed, the density contrast was higher for models with higher $\nth$ due to more shrinkage in the direction of filament crushing (Figure~\ref{fig:15}).
However, this contributes little to fragmentation.

\section{Conclusions} \label{sec:sum}

In this paper, we have performed a suite of $42$ simulations of the first star formation in a $\Lambda$CDM universe but under different initial streaming velocities (SVs).
We continue our simulation over 100,000 years after the formation of the first dense cloud, intending to study the final fate of the massive gravitationally unstable cloud hosted by a massive halo.
Our principal results are as follows:
\begin{enumerate}
\item
As initial SV increases ($v_{\rm SV} / \sigma_{\rm SV}^{\rm rec} = 0 \to 3$), the halo forms later ($dz \sim 10$) and its mass increases ($\Delta M_{\rm v} \sim 100$).
The mass of a gravitationally contracting gas inside a halo when it becomes gravitationally unstable for the first time also increases ($\Delta M_{\rm J,whole} \sim 40$).
On the other hand, the total mass of dense cores 
However, the mass of the dense core did not necessarily increase, resulting in a decrease in star formation efficiency ($\Delta(M_{\rm core,tot}/M_{\rm v,b}) \sim 10^{-2}$).
\item 
As initial SV increases, the more complex the morphology of the massive gas cloud that forms in the halo: spherical, filamentary, and complex.
The degree of temperature drop during contraction determines the morphological change of the massive gas cloud.
When the maximum to minimum temperature ratio during contraction exceeds $T_{\rm max}/T_{\rm min} \sim 10$, the shape changes from symmetric to asymmetric.
\item 
As initial SV increases, the more easily the asymmetric massive gas cloud fragment, forming an association of dense clouds.
Among $42$ models in this paper, we confirmed the simultaneous formation of up to four dense gas clouds.
Their total mass is about $2254\,\msun$, corresponding to the upper limit of stellar mass among examined models.
\end{enumerate}

\section*{Acknowledgements}

We want to thank Kei Kanaoka for his contributions to the early stages of this study.
Numerical computations were carried out on Cray XC50 at CfCA in National Astronomical Observatory of Japan and Yukawa-21 at YITP in Kyoto University.
Numerical analyses were in part carried out on the analysis servers at CfCA in National Astronomical Observatory of Japan.
This work was supported by JSPS KAKENHI Grant Numbers JP18H05222, JP21K13960, JP22H01259 (S.H.) and JP21H01123  (H.U. and S.H.), Qdai-jump Research Program 02217 (S.H.), and MEXT as ``Program for Promoting Researches on the Supercomputer Fugaku'' (Structure and Evolution of the Universe Unraveled by Fusion of Simulation and AI; Grant Number JPMXP1020230406, Project ID hp230204) (S.H.).

\section*{Data Availability}

The data underlying this article will be shared on reasonable request to the corresponding author.


\bibliographystyle{mnras}
\bibliography{ms} 

\begin{thebibliography}{}
\makeatletter
\relax
\def\mn@urlcharsother{\let\do\@makeother \do\$\do\&\do\#\do\^\do\_\do\%\do\~}
\def\mn@doi{\begingroup\mn@urlcharsother \@ifnextchar [ {\mn@doi@}
  {\mn@doi@[]}}
\def\mn@doi@[#1]#2{\def\@tempa{#1}\ifx\@tempa\@empty \href
  {http://dx.doi.org/#2} {doi:#2}\else \href {http://dx.doi.org/#2} {#1}\fi
  \endgroup}
\def\mn@eprint#1#2{\mn@eprint@#1:#2::\@nil}
\def\mn@eprint@arXiv#1{\href {http://arxiv.org/abs/#1} {{\tt arXiv:#1}}}
\def\mn@eprint@dblp#1{\href {http://dblp.uni-trier.de/rec/bibtex/#1.xml}
  {dblp:#1}}
\def\mn@eprint@#1:#2:#3:#4\@nil{\def\@tempa {#1}\def\@tempb {#2}\def\@tempc
  {#3}\ifx \@tempc \@empty \let \@tempc \@tempb \let \@tempb \@tempa \fi \ifx
  \@tempb \@empty \def\@tempb {arXiv}\fi \@ifundefined
  {mn@eprint@\@tempb}{\@tempb:\@tempc}{\expandafter \expandafter \csname
  mn@eprint@\@tempb\endcsname \expandafter{\@tempc}}}

\bibitem[\protect\citeauthoryear{{Abel}, {Bryan}  \& {Norman}}{{Abel}
  et~al.}{2002}]{Abel2002}
{Abel} T.,  {Bryan} G.~L.,   {Norman} M.~L.,  2002, \mn@doi [Science]
  {10.1126/science.295.5552.93}, \href
  {https://ui.adsabs.harvard.edu/abs/2002Sci...295...93A} {295, 93}

\bibitem[\protect\citeauthoryear{{Bonnor}}{{Bonnor}}{1956}]{Bonnor1956}
{Bonnor} W.~B.,  1956, \mn@doi [\mnras] {10.1093/mnras/116.3.351}, \href
  {https://ui.adsabs.harvard.edu/abs/1956MNRAS.116..351B} {116, 351}

\bibitem[\protect\citeauthoryear{{Bromm}, {Coppi}  \& {Larson}}{{Bromm}
  et~al.}{2002}]{Bromm2002}
{Bromm} V.,  {Coppi} P.~S.,   {Larson} R.~B.,  2002, \mn@doi [\apj]
  {10.1086/323947}, \href
  {https://ui.adsabs.harvard.edu/abs/2002ApJ...564...23B} {564, 23}

\bibitem[\protect\citeauthoryear{{Chiaki} \& {Yoshida}}{{Chiaki} \&
  {Yoshida}}{2015}]{ChiakiYoshida2015}
{Chiaki} G.,  {Yoshida} N.,  2015, \mn@doi [\mnras] {10.1093/mnras/stv1227},
  \href {https://ui.adsabs.harvard.edu/abs/2015MNRAS.451.3955C} {451, 3955}

\bibitem[\protect\citeauthoryear{{Chiou}, {Naoz}, {Burkhart}, {Marinacci}  \&
  {Vogelsberger}}{{Chiou} et~al.}{2019}]{Chiou2019}
{Chiou} Y.~S.,  {Naoz} S.,  {Burkhart} B.,  {Marinacci} F.,   {Vogelsberger}
  M.,  2019, \mn@doi [\apjl] {10.3847/2041-8213/ab263a}, \href
  {https://ui.adsabs.harvard.edu/abs/2019ApJ...878L..23C} {878, L23}

\bibitem[\protect\citeauthoryear{{Ebert}}{{Ebert}}{1955}]{Ebert1955}
{Ebert} R.,  1955, \zap, \href
  {https://ui.adsabs.harvard.edu/abs/1955ZA.....37..217E} {37, 217}

\bibitem[\protect\citeauthoryear{{Forrey}}{{Forrey}}{2013a}]{Forrey2013b}
{Forrey} R.~C.,  2013a, \mn@doi [\pra] {10.1103/PhysRevA.88.052709}, \href
  {https://ui.adsabs.harvard.edu/abs/2013PhRvA..88e2709F} {88, 052709}

\bibitem[\protect\citeauthoryear{{Forrey}}{{Forrey}}{2013b}]{Forrey2013a}
{Forrey} R.~C.,  2013b, \mn@doi [\apjl] {10.1088/2041-8205/773/2/L25}, \href
  {https://ui.adsabs.harvard.edu/abs/2013ApJ...773L..25F} {773, L25}

\bibitem[\protect\citeauthoryear{{Galli} \& {Palla}}{{Galli} \&
  {Palla}}{2013}]{GalliPalla2013}
{Galli} D.,  {Palla} F.,  2013, \mn@doi [\araa]
  {10.1146/annurev-astro-082812-141029}, \href
  {https://ui.adsabs.harvard.edu/abs/2013ARA&A..51..163G} {51, 163}

\bibitem[\protect\citeauthoryear{{Greif}, {White}, {Klessen}  \&
  {Springel}}{{Greif} et~al.}{2011}]{Greif2011}
{Greif} T.~H.,  {White} S. D.~M.,  {Klessen} R.~S.,   {Springel} V.,  2011,
  \mn@doi [\apj] {10.1088/0004-637X/736/2/147}, \href
  {https://ui.adsabs.harvard.edu/abs/2011ApJ...736..147G} {736, 147}

\bibitem[\protect\citeauthoryear{{Hahn} \& {Abel}}{{Hahn} \&
  {Abel}}{2011}]{Hahn2011}
{Hahn} O.,  {Abel} T.,  2011, \mn@doi [\mnras]
  {10.1111/j.1365-2966.2011.18820.x}, \href
  {https://ui.adsabs.harvard.edu/abs/2011MNRAS.415.2101H} {415, 2101}

\bibitem[\protect\citeauthoryear{{Hirano} \& {Bromm}}{{Hirano} \&
  {Bromm}}{2017}]{Hirano2017b}
{Hirano} S.,  {Bromm} V.,  2017, \mn@doi [\mnras] {10.1093/mnras/stx1220},
  \href {https://ui.adsabs.harvard.edu/abs/2017MNRAS.470..898H} {470, 898}

\bibitem[\protect\citeauthoryear{{Hirano}, {Hosokawa}, {Yoshida}, {Umeda},
  {Omukai}, {Chiaki}  \& {Yorke}}{{Hirano} et~al.}{2014}]{Hirano2014}
{Hirano} S.,  {Hosokawa} T.,  {Yoshida} N.,  {Umeda} H.,  {Omukai} K.,
  {Chiaki} G.,   {Yorke} H.~W.,  2014, \mn@doi [\apj]
  {10.1088/0004-637X/781/2/60}, \href
  {https://ui.adsabs.harvard.edu/abs/2014ApJ...781...60H} {781, 60}

\bibitem[\protect\citeauthoryear{{Hirano}, {Hosokawa}, {Yoshida}, {Omukai}  \&
  {Yorke}}{{Hirano} et~al.}{2015}]{Hirano2015}
{Hirano} S.,  {Hosokawa} T.,  {Yoshida} N.,  {Omukai} K.,   {Yorke} H.~W.,
  2015, \mn@doi [\mnras] {10.1093/mnras/stv044}, \href
  {https://ui.adsabs.harvard.edu/abs/2015MNRAS.448..568H} {448, 568}

\bibitem[\protect\citeauthoryear{{Hirano}, {Hosokawa}, {Yoshida}  \&
  {Kuiper}}{{Hirano} et~al.}{2017}]{Hirano2017a}
{Hirano} S.,  {Hosokawa} T.,  {Yoshida} N.,   {Kuiper} R.,  2017, \mn@doi
  [Science] {10.1126/science.aai9119}, \href
  {https://ui.adsabs.harvard.edu/abs/2017Sci...357.1375H} {357, 1375}

\bibitem[\protect\citeauthoryear{{Hirano}, {Yoshida}, {Sakurai}  \&
  {Fujii}}{{Hirano} et~al.}{2018}]{Hirano2018}
{Hirano} S.,  {Yoshida} N.,  {Sakurai} Y.,   {Fujii} M.~S.,  2018, \mn@doi
  [\apj] {10.3847/1538-4357/aaaaba}, \href
  {https://ui.adsabs.harvard.edu/abs/2018ApJ...855...17H} {855, 17}

\bibitem[\protect\citeauthoryear{{Hosokawa}, {Omukai}, {Yoshida}  \&
  {Yorke}}{{Hosokawa} et~al.}{2011}]{Hosokawa2011}
{Hosokawa} T.,  {Omukai} K.,  {Yoshida} N.,   {Yorke} H.~W.,  2011, \mn@doi
  [Science] {10.1126/science.1207433}, \href
  {https://ui.adsabs.harvard.edu/abs/2011Sci...334.1250H} {334, 1250}

\bibitem[\protect\citeauthoryear{{Hosokawa}, {Omukai}  \& {Yorke}}{{Hosokawa}
  et~al.}{2012}]{Hosokawa2012a}
{Hosokawa} T.,  {Omukai} K.,   {Yorke} H.~W.,  2012, \mn@doi [\apj]
  {10.1088/0004-637X/756/1/93}, \href
  {https://ui.adsabs.harvard.edu/abs/2012ApJ...756...93H} {756, 93}

\bibitem[\protect\citeauthoryear{{Hosokawa}, {Hirano}, {Kuiper}, {Yorke},
  {Omukai}  \& {Yoshida}}{{Hosokawa} et~al.}{2016}]{Hosokawa2016}
{Hosokawa} T.,  {Hirano} S.,  {Kuiper} R.,  {Yorke} H.~W.,  {Omukai} K.,
  {Yoshida} N.,  2016, \mn@doi [\apj] {10.3847/0004-637X/824/2/119}, \href
  {https://ui.adsabs.harvard.edu/abs/2016ApJ...824..119H} {824, 119}

\bibitem[\protect\citeauthoryear{{Hummel}, {Stacy}, {Jeon}, {Oliveri}  \&
  {Bromm}}{{Hummel} et~al.}{2015}]{Hummel2005}
{Hummel} J.~A.,  {Stacy} A.,  {Jeon} M.,  {Oliveri} A.,   {Bromm} V.,  2015,
  \mn@doi [\mnras] {10.1093/mnras/stv1902}, \href
  {https://ui.adsabs.harvard.edu/abs/2015MNRAS.453.4136H} {453, 4136}

\bibitem[\protect\citeauthoryear{{Inayoshi}, {Visbal}  \&
  {Kashiyama}}{{Inayoshi} et~al.}{2015}]{Inayoshi2015}
{Inayoshi} K.,  {Visbal} E.,   {Kashiyama} K.,  2015, \mn@doi [\mnras]
  {10.1093/mnras/stv1654}, \href
  {https://ui.adsabs.harvard.edu/abs/2015MNRAS.453.1692I} {453, 1692}

\bibitem[\protect\citeauthoryear{{Inayoshi}, {Visbal}  \& {Haiman}}{{Inayoshi}
  et~al.}{2020}]{Inayoshi2020}
{Inayoshi} K.,  {Visbal} E.,   {Haiman} Z.,  2020, \mn@doi [\araa]
  {10.1146/annurev-astro-120419-014455}, \href
  {https://ui.adsabs.harvard.edu/abs/2020ARA&A..58...27I} {58, 27}

\bibitem[\protect\citeauthoryear{{Inutsuka} \& {Miyama}}{{Inutsuka} \&
  {Miyama}}{1992}]{Inutsuka1992}
{Inutsuka} S.-I.,  {Miyama} S.~M.,  1992, \mn@doi [\apj] {10.1086/171162},
  \href {https://ui.adsabs.harvard.edu/abs/1992ApJ...388..392I} {388, 392}

\bibitem[\protect\citeauthoryear{{Inutsuka} \& {Miyama}}{{Inutsuka} \&
  {Miyama}}{1997}]{Inutsuka1997}
{Inutsuka} S.-i.,  {Miyama} S.~M.,  1997, \mn@doi [\apj] {10.1086/303982},
  \href {https://ui.adsabs.harvard.edu/abs/1997ApJ...480..681I} {480, 681}

\bibitem[\protect\citeauthoryear{{Kitsionas} \& {Whitworth}}{{Kitsionas} \&
  {Whitworth}}{2002}]{Kitsionas2002}
{Kitsionas} S.,  {Whitworth} A.~P.,  2002, \mn@doi [\mnras]
  {10.1046/j.1365-8711.2002.05115.x}, \href
  {https://ui.adsabs.harvard.edu/abs/2002MNRAS.330..129K} {330, 129}

\bibitem[\protect\citeauthoryear{{Klessen} \& {Glover}}{{Klessen} \&
  {Glover}}{2023}]{Klessen2023}
{Klessen} R.~S.,  {Glover} S. C.~O.,  2023, \mn@doi [\araa]
  {10.1146/annurev-astro-071221-053453}, \href
  {https://ui.adsabs.harvard.edu/abs/2023ARA&A..61...65K} {61, 65}

\bibitem[\protect\citeauthoryear{{Kulkarni}, {Visbal}  \& {Bryan}}{{Kulkarni}
  et~al.}{2021}]{Kulkarni2021}
{Kulkarni} M.,  {Visbal} E.,   {Bryan} G.~L.,  2021, \mn@doi [\apj]
  {10.3847/1538-4357/ac08a3}, \href
  {https://ui.adsabs.harvard.edu/abs/2021ApJ...917...40K} {917, 40}

\bibitem[\protect\citeauthoryear{{Lake} et~al.,}{{Lake}
  et~al.}{2023}]{Lake2023}
{Lake} W.,  et~al., 2023, \mn@doi [arXiv e-prints] {10.48550/arXiv.2306.01047},
  \href {https://ui.adsabs.harvard.edu/abs/2023arXiv230601047L} {p.
  arXiv:2306.01047}

\bibitem[\protect\citeauthoryear{{Latif}, {Schleicher}, {Schmidt}  \&
  {Niemeyer}}{{Latif} et~al.}{2013}]{Latif2013}
{Latif} M.~A.,  {Schleicher} D.~R.~G.,  {Schmidt} W.,   {Niemeyer} J.,  2013,
  \mn@doi [\mnras] {10.1093/mnras/stt834}, \href
  {https://ui.adsabs.harvard.edu/abs/2013MNRAS.433.1607L} {433, 1607}

\bibitem[\protect\citeauthoryear{{McKee} \& {Tan}}{{McKee} \&
  {Tan}}{2008}]{McKee2008}
{McKee} C.~F.,  {Tan} J.~C.,  2008, \mn@doi [\apj] {10.1086/587434}, \href
  {https://ui.adsabs.harvard.edu/abs/2008ApJ...681..771M} {681, 771}

\bibitem[\protect\citeauthoryear{{Naoz} \& {Narayan}}{{Naoz} \&
  {Narayan}}{2014}]{Naoz2014}
{Naoz} S.,  {Narayan} R.,  2014, \mn@doi [\apjl] {10.1088/2041-8205/791/1/L8},
  \href {https://ui.adsabs.harvard.edu/abs/2014ApJ...791L...8N} {791, L8}

\bibitem[\protect\citeauthoryear{{Omukai}}{{Omukai}}{2001}]{Omukai2001}
{Omukai} K.,  2001, \mn@doi [\apj] {10.1086/318296}, \href
  {https://ui.adsabs.harvard.edu/abs/2001ApJ...546..635O} {546, 635}

\bibitem[\protect\citeauthoryear{{Omukai} \& {Nishi}}{{Omukai} \&
  {Nishi}}{1998}]{Omukai1998}
{Omukai} K.,  {Nishi} R.,  1998, \mn@doi [\apj] {10.1086/306395}, \href
  {https://ui.adsabs.harvard.edu/abs/1998ApJ...508..141O} {508, 141}

\bibitem[\protect\citeauthoryear{{Park}, {Ricotti}  \& {Sugimura}}{{Park}
  et~al.}{2021}]{Park2021}
{Park} J.,  {Ricotti} M.,   {Sugimura} K.,  2021, \mn@doi [\mnras]
  {10.1093/mnras/stab2999}, \href
  {https://ui.adsabs.harvard.edu/abs/2021MNRAS.508.6176P} {508, 6176}

\bibitem[\protect\citeauthoryear{{Popa}, {Naoz}, {Marinacci}  \&
  {Vogelsberger}}{{Popa} et~al.}{2016}]{Popa2016}
{Popa} C.,  {Naoz} S.,  {Marinacci} F.,   {Vogelsberger} M.,  2016, \mn@doi
  [\mnras] {10.1093/mnras/stw1045}, \href
  {https://ui.adsabs.harvard.edu/abs/2016MNRAS.460.1625P} {460, 1625}

\bibitem[\protect\citeauthoryear{{Regan}, {Wise}, {Woods}, {Downes}, {O'Shea}
  \& {Norman}}{{Regan} et~al.}{2020}]{Regan2020}
{Regan} J.~A.,  {Wise} J.~H.,  {Woods} T.~E.,  {Downes} T.~P.,  {O'Shea} B.~W.,
    {Norman} M.~L.,  2020, \mn@doi [The Open Journal of Astrophysics]
  {10.21105/astro.2008.08090}, \href
  {https://ui.adsabs.harvard.edu/abs/2020OJAp....3E..15R} {3, 15}

\bibitem[\protect\citeauthoryear{{Ripamonti}}{{Ripamonti}}{2007}]{Ripamonti2007}
{Ripamonti} E.,  2007, \mn@doi [\mnras] {10.1111/j.1365-2966.2007.11460.x},
  \href {https://ui.adsabs.harvard.edu/abs/2007MNRAS.376..709R} {376, 709}

\bibitem[\protect\citeauthoryear{{Schauer}, {Glover}, {Klessen}  \&
  {Clark}}{{Schauer} et~al.}{2021}]{Schauer2021}
{Schauer} A. T.~P.,  {Glover} S. C.~O.,  {Klessen} R.~S.,   {Clark} P.,  2021,
  \mn@doi [\mnras] {10.1093/mnras/stab1953}, \href
  {https://ui.adsabs.harvard.edu/abs/2021MNRAS.507.1775S} {507, 1775}

\bibitem[\protect\citeauthoryear{{Springel}}{{Springel}}{2005}]{Springel2005}
{Springel} V.,  2005, \mn@doi [\mnras] {10.1111/j.1365-2966.2005.09655.x},
  \href {https://ui.adsabs.harvard.edu/abs/2005MNRAS.364.1105S} {364, 1105}

\bibitem[\protect\citeauthoryear{{Stacy} \& {Bromm}}{{Stacy} \&
  {Bromm}}{2013}]{Stacy2013}
{Stacy} A.,  {Bromm} V.,  2013, \mn@doi [\mnras] {10.1093/mnras/stt789}, \href
  {https://ui.adsabs.harvard.edu/abs/2013MNRAS.433.1094S} {433, 1094}

\bibitem[\protect\citeauthoryear{{Stacy}, {Bromm}  \& {Loeb}}{{Stacy}
  et~al.}{2011}]{Stacy2011}
{Stacy} A.,  {Bromm} V.,   {Loeb} A.,  2011, \mn@doi [\apjl]
  {10.1088/2041-8205/730/1/L1}, \href
  {https://ui.adsabs.harvard.edu/abs/2011ApJ...730L...1S} {730, L1}

\bibitem[\protect\citeauthoryear{{Stacy}, {Bromm}  \& {Lee}}{{Stacy}
  et~al.}{2016}]{Stacy2016}
{Stacy} A.,  {Bromm} V.,   {Lee} A.~T.,  2016, \mn@doi [\mnras]
  {10.1093/mnras/stw1728}, \href
  {https://ui.adsabs.harvard.edu/abs/2016MNRAS.462.1307S} {462, 1307}

\bibitem[\protect\citeauthoryear{{Sugimura}, {Matsumoto}, {Hosokawa}, {Hirano}
  \& {Omukai}}{{Sugimura} et~al.}{2020}]{Sugimura2020}
{Sugimura} K.,  {Matsumoto} T.,  {Hosokawa} T.,  {Hirano} S.,   {Omukai} K.,
  2020, \mn@doi [\apjl] {10.3847/2041-8213/ab7d37}, \href
  {https://ui.adsabs.harvard.edu/abs/2020ApJ...892L..14S} {892, L14}

\bibitem[\protect\citeauthoryear{{Susa}}{{Susa}}{2013}]{Susa2013}
{Susa} H.,  2013, \mn@doi [\apj] {10.1088/0004-637X/773/2/185}, \href
  {https://ui.adsabs.harvard.edu/abs/2013ApJ...773..185S} {773, 185}

\bibitem[\protect\citeauthoryear{{Susa}}{{Susa}}{2019}]{susa2019}
{Susa} H.,  2019, \mn@doi [\apj] {10.3847/1538-4357/ab1b6f}, \href
  {https://ui.adsabs.harvard.edu/abs/2019ApJ...877...99S} {877, 99}

\bibitem[\protect\citeauthoryear{{Susa}, {Hasegawa}  \& {Tominaga}}{{Susa}
  et~al.}{2014}]{Susa2014}
{Susa} H.,  {Hasegawa} K.,   {Tominaga} N.,  2014, \mn@doi [\apj]
  {10.1088/0004-637X/792/1/32}, \href
  {https://ui.adsabs.harvard.edu/abs/2014ApJ...792...32S} {792, 32}

\bibitem[\protect\citeauthoryear{{Tegmark}, {Silk}, {Rees}, {Blanchard}, {Abel}
   \& {Palla}}{{Tegmark} et~al.}{1997}]{Tegmark1997}
{Tegmark} M.,  {Silk} J.,  {Rees} M.~J.,  {Blanchard} A.,  {Abel} T.,   {Palla}
  F.,  1997, \mn@doi [\apj] {10.1086/303434}, \href
  {https://ui.adsabs.harvard.edu/abs/1997ApJ...474....1T} {474, 1}

\bibitem[\protect\citeauthoryear{{Tseliakhovich} \& {Hirata}}{{Tseliakhovich}
  \& {Hirata}}{2010}]{Tseliakhovich2010}
{Tseliakhovich} D.,  {Hirata} C.,  2010, \mn@doi [\prd]
  {10.1103/PhysRevD.82.083520}, \href
  {https://ui.adsabs.harvard.edu/abs/2010PhRvD..82h3520T} {82, 083520}

\bibitem[\protect\citeauthoryear{{Tseliakhovich}, {Barkana}  \&
  {Hirata}}{{Tseliakhovich} et~al.}{2011}]{Tseliakhovich2011}
{Tseliakhovich} D.,  {Barkana} R.,   {Hirata} C.~M.,  2011, \mn@doi [\mnras]
  {10.1111/j.1365-2966.2011.19541.x}, \href
  {https://ui.adsabs.harvard.edu/abs/2011MNRAS.418..906T} {418, 906}

\bibitem[\protect\citeauthoryear{{Uysal} \& {Hartwig}}{{Uysal} \&
  {Hartwig}}{2023}]{UysalHartwig2023}
{Uysal} B.,  {Hartwig} T.,  2023, \mn@doi [\mnras] {10.1093/mnras/stad350},
  \href {https://ui.adsabs.harvard.edu/abs/2023MNRAS.520.3229U} {520, 3229}

\bibitem[\protect\citeauthoryear{{Wise}, {Regan}, {O'Shea}, {Norman}, {Downes}
  \& {Xu}}{{Wise} et~al.}{2019}]{Wise2019}
{Wise} J.~H.,  {Regan} J.~A.,  {O'Shea} B.~W.,  {Norman} M.~L.,  {Downes}
  T.~P.,   {Xu} H.,  2019, \mn@doi [\nat] {10.1038/s41586-019-0873-4}, \href
  {https://ui.adsabs.harvard.edu/abs/2019Natur.566...85W} {566, 85}

\bibitem[\protect\citeauthoryear{{Yoshida}, {Sokasian}, {Hernquist}  \&
  {Springel}}{{Yoshida} et~al.}{2003}]{Yoshida2003}
{Yoshida} N.,  {Sokasian} A.,  {Hernquist} L.,   {Springel} V.,  2003, \mn@doi
  [\apj] {10.1086/378852}, \href
  {https://ui.adsabs.harvard.edu/abs/2003ApJ...598...73Y} {598, 73}

\bibitem[\protect\citeauthoryear{{Yoshida}, {Oh}, {Kitayama}  \&
  {Hernquist}}{{Yoshida} et~al.}{2007}]{Yoshida2007}
{Yoshida} N.,  {Oh} S.~P.,  {Kitayama} T.,   {Hernquist} L.,  2007, \mn@doi
  [\apj] {10.1086/518227}, \href
  {https://ui.adsabs.harvard.edu/abs/2007ApJ...663..687Y} {663, 687}

\bibitem[\protect\citeauthoryear{{Yoshida}, {Omukai}  \& {Hernquist}}{{Yoshida}
  et~al.}{2008}]{Yoshida2008}
{Yoshida} N.,  {Omukai} K.,   {Hernquist} L.,  2008, \mn@doi [Science]
  {10.1126/science.1160259}, \href
  {https://ui.adsabs.harvard.edu/abs/2008Sci...321..669Y} {321, 669}

\makeatother
\end{thebibliography}






\bsp	
\label{lastpage}
\end{document}